\documentclass[preprint2]{emulateapj}

\usepackage{graphicx}
\usepackage{float}
\usepackage{bm} 
\usepackage{amssymb}
\usepackage{amsmath} 
\usepackage[export]{adjustbox}
\usepackage{mathrsfs} 
\usepackage{mathtools}

\def\rt{r_{\rm t}}

\shorttitle{Model of the `dog leg' stream}
\shortauthors{Amorisco et al.}

\begin{document}


\title{A dwarf galaxy's transformation and a massive galaxy's edge: autopsy of kill and killer in NGC 1097 }


\author{N. C. Amorisco\altaffilmark{1}}
\affil{Dark Cosmology Centre, Niels Bohr Institute, University of Copenhagen, Juliane Maries Vej 30, DK-2100 Copenhagen, Denmark}

\author{D. Martinez-Delgado}
\affil{Astronomisches Rechen-Institut, Zentrum f{\"u}r Astronomie der Universit{\"a}t Heidelberg, M{\"o}nchhofstr. 12-14, 69120 Heidelberg, Germany}

\author{J. Schedler}
\affil{CHART32, CTIO, Chile}




\altaffiltext{1}{amorisco@dark-cosmology.dk}


\begin{abstract}
We present a detailed dynamical analysis of the thin, extended stellar stream that encircles the disk galaxy NGC 1097.
Within a fully statistical framework, we model the surface brightness map of the so called `dog leg stream', using mock streams
generated as in \citet{NA15}, and new deep imaging data obtained with the CHART32 telescope for the Stellar Tidal Stream Survey.
We reconstruct the post-infall evolution of the progenitor galaxy, which, we find, has experienced three pericentric passages, and lost more than two orders 
of magnitude in mass. At infall, $5.4\pm0.6$ Gyr ago, 
the progenitor was a disky dwarf with a mass of $\log_{10}(m/ M_\odot)=10.35\pm0.25$ within its tidal radius, $r_{\rm t}=3.4\pm1$ kpc. 
We illustrate how the sharp 90 degree turn in the stream, identifying the `dog leg', is the signature of the progenitor's internal rotation, 
inclined with respect to the orbital plane and imprinted in the escape conditions of the stream members.
Today, the remnant is observed to be a nucleated 
dwarf galaxy, with a line-of-sight velocity of $v_{\rm p, los}^{\rm obs}=-30\pm 30$ kms$^{-1}$ with respect to NGC 1097, and 
a total luminosity of $3.3\times 10^7 L_{V,\odot}$ \citep{PG10}.
Our independent analysis predicts a line-of-sight velocity of $v_{\rm p, los}=-51^{-17}_{+14}$ kms$^{-1}$ and measures its current total mass at 
$\log_{10}(m/ M_\odot)=7.4^{+0.6}_{-0.8}$, implying that the compact nucleus \citep[$L\approx 6.9\times 10^5 L_{V,\odot}$, ][]{PG10}  
is soon destined to become a low-luminosity Ultra Compact Dwarf galaxy.
The progenitor's orbit is strongly radial, with a pericenter of a few kpc, and 
an apocenter reaching $r_{\rm apo}=150\pm 15$ kpc, more than a half of the inferred virial radius of the host, $r_{200}=250\pm20$ kpc. We find 
that NGC 1097 has a mass of $M_{200}=1.8^{+0.5}_{-0.4} \times 10^{12}\; M_{\odot}$, and its inferred concentration $c_{200}=6.7^{+2.4}_{-1.3}$ is in very 
good agreement with the expectation of the cold dark matter model. We can describe the stream almost down to the noise using a spherical host potential, and 
find that this would not be possible in case of a halo that is substantially triaxial at large radii. In the morphology of the stream, we can see the logarithmic density slope of the 
total density profile, $\gamma$, bending from its inner value, $\gamma(r_{\rm peri})=1.5\pm0.15$, and steepening towards large radii. 
For the first time on an individual extragalactic halo, we measure the outer slope of the density profile, $\gamma(0.6r_{200})=3.9\pm0.5$.
This demonstrates the promise of the newborn field of detailed, statistical modelling of extragalactic tidal streams.
\end{abstract}



\section{Introduction}

Recent deep wide-field imaging surveys of stellar tidal streams have mainly focused on nearby Milky Way-like 
spiral galaxies that were suspected to contain diffuse-light, very faint features in their outer regions  based on 
existing data from large scale surveys, such as POSSII \citep{MH97} and the SDSS \citep{Mi11}, 
or amateur images \citep{MD08,MD10}. The morphology of the detected tidal streams include 
great circle streams resembling the Sagittarius stream around our Galaxy, isolated shells, giant 
clouds of debris floating within galactic halos, spike-like features emerging from galactic disks, and 
large-scale diffuse structures that may be related to the remnants of ancient, already thoroughly 
disrupted satellites \citep{MD10}. This extraordinary variety of morphologies represents 
one of the first comprehensive pieces of evidence for the hierarchical formation scenarios predicted 
by cosmological models \citep[e.g.,][]{SW91,JB05, JBS08, AC10}. Encouraged by these results, the Stellar Tidal Stream 
Survey (PI. Mart«õnez-Delgado) is carrying out the first systematic survey of stellar tidal streams to a surface 
brightness sensitivity of $\mu_{r}$=28.5 mag arcsec$^{-2}$ using a network of small, robotic telescopes 
placed on different continents.

These debris contain precious information on both (i) the evolutionary struggles of the progenitor dwarf galaxies, transformed and partially 
destroyed by the tides; (ii) the formation history of the massive host, its halo populations and the structure of its dark matter halo. 
However, rewinding the movie of these accretion events has so far been difficult. It is often possible to 
identify a set of properties of progenitor and host that provides a qualitative resemblance to the observed
stream \citep{MD08, MD09, AC11, CF14, MD14}, but a full exploration of the wide parameter space is exceedingly 
expensive to perform with N-body simulations, meaning that the extraction of reliable measurements is not possible. 
Therefore, it remains extremely difficult to provide robust confidence limits on several interesting 
quantities, such as the progenitor's initial mass, infall time and orbital properties, as well 
as host halo mass, concentration and detailed density profile.

Extended and reasonably cold streams represent one of the least degenerate probes of the gravitational
potential in which they orbit, as their global morphology directly traces the density slope and geometrical structure of the host galaxy. 
While it is not correct to assume that a stream traces an orbit \citep{JS13} as different positions along its length are 
directly associated to a gradient in orbital energy \citep[][]{KJ01, NA15}, it is qualitatively correct to say that, as for orbits,
different density profiles imply substantially different shapes. For example, the opening angle 
between successive apocenters \citep{VB14} is certainly a function of the underlying gravitational potential.
Furthermore, in addition to what is easily understood within the single orbit interpretation,
streams also provide additional constraining power, in the shifting 
galactocentric distances of successive apocenters (increasing along the trailing tail and shrinking along the leading tail), 
as well as the links between the stream's total length and width and the inner density slope of the host \citep{NA15}.

The interplay between the central dark matter distribution and the lifecycle of baryons 
is one of the crucial ingredients of the galaxy formation recipe. The predictions of dark-matter-only cosmological 
simulations are quite clear in this respect \citep[see e.g., ][]{JN97, JN04}, but we still lack a thorough understanding of the equilibrium between competing 
processes such as adiabatic contraction \citep{Bl86, OG04} and stellar feedback \citep{JN96, Ma06, AP12} across the galactic mass scale. 
Therefore, the inner total and dark matter density slopes are currently subject to close observational attention, mainly through stellar and/or gaseous kinematics 
\citep[e.g.][and references therein]{deB10, MC13} and strong lensing analyses \citep[e.g.,][]{LK09, AS12, CG12}.

\begin{figure}
\includegraphics[width=\columnwidth]{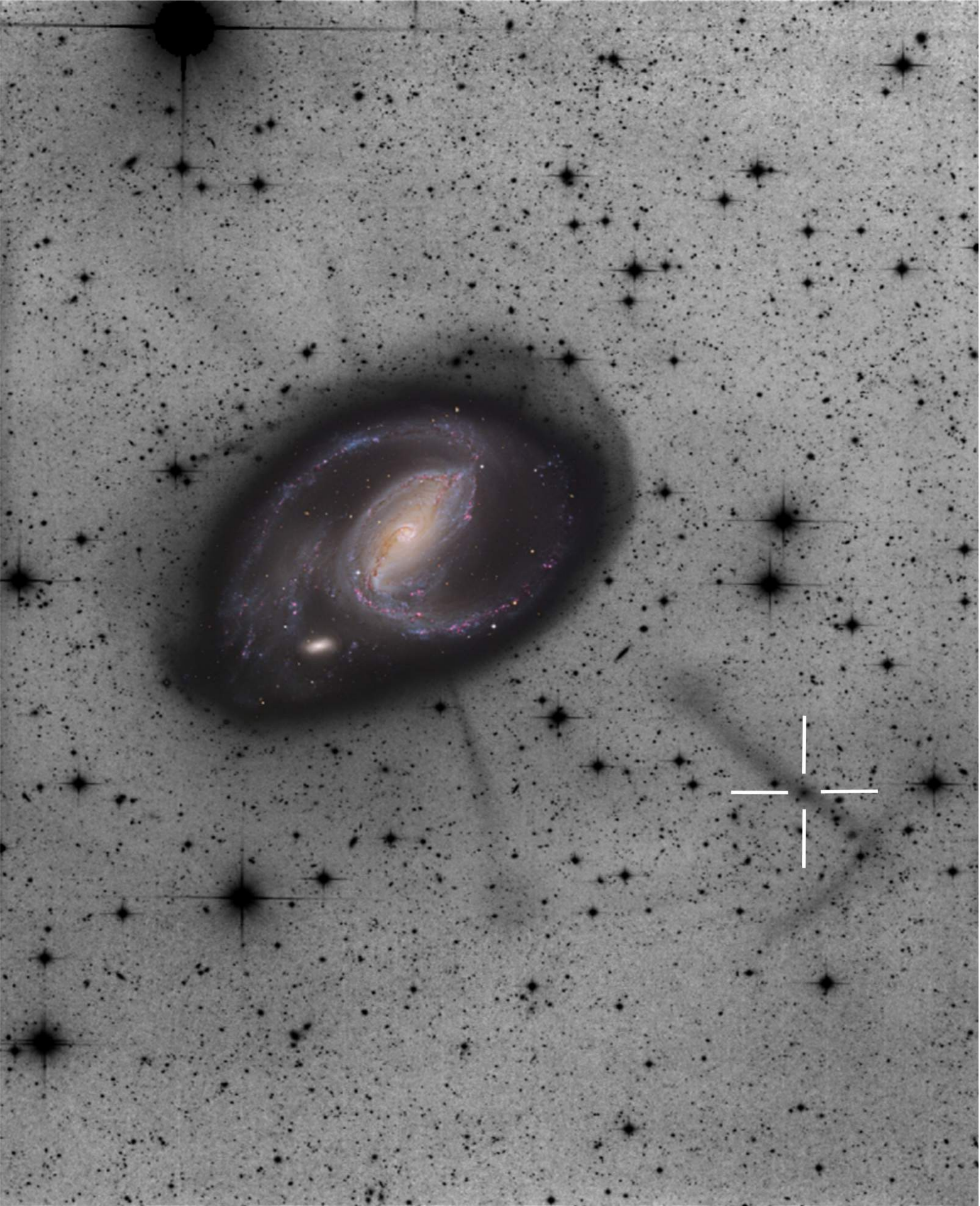}
\caption{Luminance filter image of the ``dog leg stream'' around NGC 1097 obtained with the CHART32 amateur 
telescope (see Sec. 2.1). The image covers a total field of 
view of 25.2 arcmin x 31.2 arcmin, that at the distance of 16.96 Mpc corresponds a projected area of 127 x 157 kpc.
The thin white cross identifies the remnant of the progenitor of the stream, currently a nucleated dwarf galaxy \citep{PG10}.
\label{rawdat}}
\end{figure}

Conversely, it is considerably more difficult to constrain the outer density profile of galactic dark matter haloes.
Deep HI rotation curves of disk galaxies extend to $\approx 10$ disk scalengths, or $\approx 50$ kpc \citep[e.g.][]{deB08}, just beginning to 
explore the regime $r\gtrsim r_0$, where $r_0$ is the scale radius of the halo. There's also good
hope for dynamical analyses based on discrete tracers in the haloes of massive galaxies: using Planetary
Nebulae and Globular Clusters it is possible to probe scales of $\approx100$ kpc, henceforth reaching a fraction 
of the virial radius \citep{NN11, VP13, JB14, Lo15}. 
However, modelling has to struggle against the intrinsic degeneracies of pressure supported
equilibria \citep[e.g.][]{JB82}, and while resorting to the joint modelling of distinct subpopulations is helpful in 
inferring the properties of the inner profile \citep{AA14, NN14}, measuring the mass and density profile at large radii
might just be too difficult.  

As a consequence, observational constraints on the outer density profile of galaxy-sized dark matter haloes
are extremely scarce. \citet{RG07} have used a combined strong and weak lensing analysis on a sample of 22 early type
massive lenses (aperture-averaged central velocity dispersions between 200 and 300 kms$^{-1}$), and inferred the average 
dark matter distribution out to $\approx300$ kpc, or $\approx0.75 r_{\rm vir}$. They find that the weak lensing signal can be 
satisfactorily described by a density profile having an outer slope of $\gamma_{o}=-3$. However, the radial range actually probed
does not extend much beyond the regime $\gamma_{DM}(r)\approx-2$, nor the available data is sufficient to allow for a comparison between
models with different asymptotic density slopes. Analogous weak lensing analyses on stacked data are also available for more massive systems, 
as galaxy groups and clusters. \citet{RM06} and \citet{KU11} can see the total density slope bending from an inner value 
compatible with a typical NFW profile, $\gamma_i=-1$, and steepening in a manner that can be well described by a profile with $\gamma_o=-3$.

Differently from the constraints we just mentioned, which use stacked data and only reveal the average properties of haloes, 
extended tidal features offer us the opportunity to explore this dark matter dominated realm in individual galaxies, without having to worry about the additional 
biases introduced by stacking and by the homology hypothesis. Detailed modelling of extragalactic tidal features
is definitely a field in its infancy, but the quick developments in the description and modelling of tidal streams \citep[see e.g.,][and references therein]{NA15, AK15},
together with the observational race towards such low surface brightness targets \citep{MD10, RA14, DP15} ensure fast and promising 
developments. Expectations also include a clearer picture of the evolution of dwarf galaxies, by the reconstruction of the 
post infall evolution of the progenitors of the studied streams. 

The extended jet-like features apparently emanating from the central regions of NGC 1097 are not a recent discovery \citep[e.g.][]{Ar76};
however, their real nature has remained elusive for almost two decades, until their possible associations with radio jets 
was finally ruled out \citep[][and references therein]{We97}. The four plumes in the X-shaped stream (see Fig.~\ref{rawdat}) 
reach out to projected distances of over 60 kpc, extending well beyond the barred disk of NGC 1097. The apparently longer plume 
displays a peculiar, sharp 90 degree turn, which has originated the name of ``dog leg stream" \citep{PG10}. 
The same authors identify a nucleated dwarf -- total luminosity $\approx3.3\times10^7 L_{V,\odot}$ 
and half light radius of $\approx 300$ pc -- well aligned along the `leg' of the stream, as indicated in Fig.~\ref{rawdat}. They spectroscopically confirm its association with NGC 1097,
and show that the stream appears to be emanating directly from this ``dwarf Spheroidal-like'' object, which is then recognised as the remnant of the progenitor.

The minor merger scenario has been explored by \citet{HW03}, which
try to roughly reproduce the stream's morphology using N-body simulations of the disruption of a disky dwarf.  
They conclude that the peculiar X-shape of the stream in NGC 1097 is due to the strong disk component of the host, 
as they do not seem to find viable models when using spherical potentials. 

In this paper, we use new deep imaging data of the stream in NGC 1097 and a new dynamical technique to reconstruct
this accretion event within a fully statistical framework. We model the surface brightness of the stream using mock streams 
orbiting static, spherical potentials \citep{NA15}. These are generated within a mixed parametric/non-parametric approach, which allows us to 
follow the mass evolution and disruption of the progenitor. This technique is computationally efficient enough to allow for a 
maximum likelihood implementation and a systematic exploration of the parameter space, so that reliable confidence intervals 
can be inferred for the properties of both progenitor and host. Section~2 presents the available data; Section~3 provides an overview 
of the generative technique; Section~4 shows that the projected shape of an orbit is sufficient to constrain the underlying 
gravitational potential; Section~5 and~6 present the results of this paper, respectively pertaining host galaxy and progenitor dwarf galaxy; 
Section~7 comments on the limits of this modelling approach and discusses our results.

\begin{figure}
\centering
\includegraphics[width=.49\columnwidth]{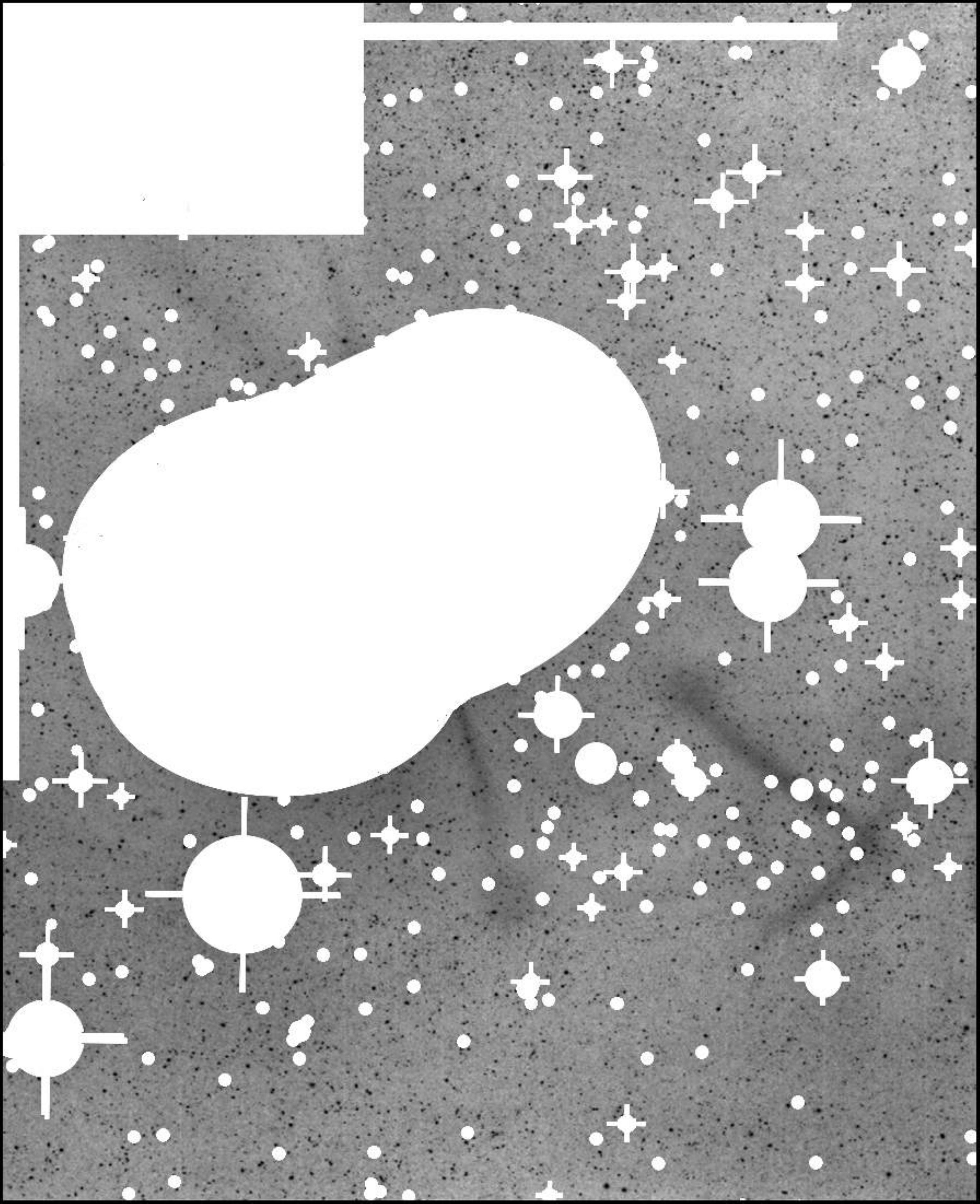}
\includegraphics[width=.49\columnwidth]{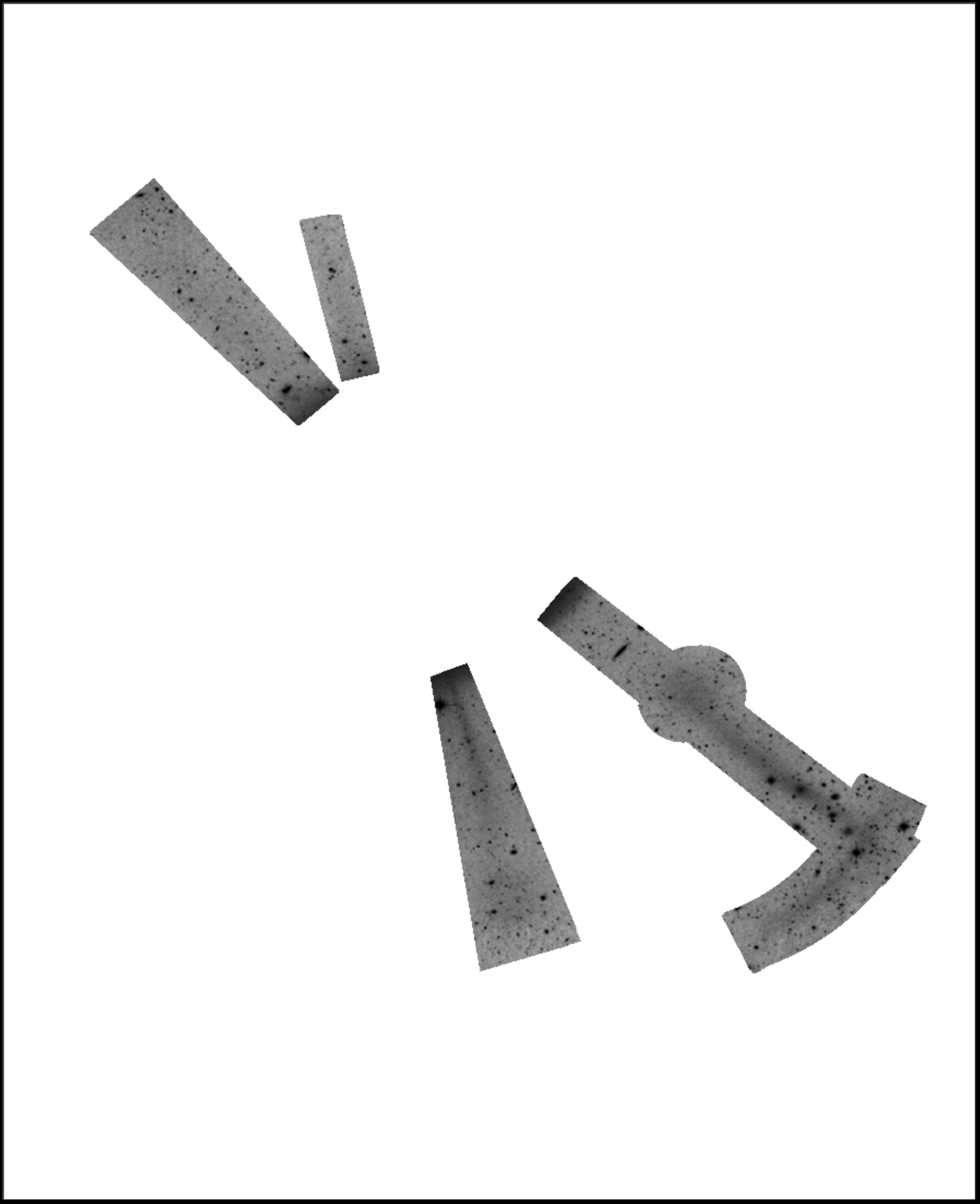}
\caption{The masks used to construct the surface brightness map of the stream. {\it Left:} 
the host galaxy, foreground stars and bright background galaxies are masked out. 
{\it Right:} the region of the stream is isolated to learn the properties of the noise in the 
remainder of the image. \label{masks}}
\end{figure}
%

\section{Observational Data}

\subsection{Observations}
Deep imaging of the field around the spiral galaxy NGC1097 was collected at the remote observatory 
CHART32\footnote{Chilean Advanced Robotic Telescope} (Cerro Tololo Interamerican Observatory,Chile)  
with a 80-cm aperture f/7 corrected Cassegrain telescope as part of the Stellar Tidal Stream Survey 
(PI. Martinez-Delgado).  A FLI PL-16803 CCD camera was used and provided a pixel scale of 0.331$^\prime$$^\prime$/px-1 
over a 22$^\prime$ $\times$ 22$^\prime$ field of view. Two overlapping sets of 20 $\times$ 1200 s individual image frames were obtained 
remotely with a Baader luminance filter over several photometric nights between September and November 2013.
Each individual exposure was reduced following standard image processing procedures for dark subtraction, 
bias correction, and flat fielding adopted for the larger stream survey \citep{MD10}. The images were combined 
to create a final co-added image with a total exposure time of 48,000 s. The resulting luminance
image is displayed in Fig.~1. A detailed description of the photometric calibration of the data, together with our 
near-infrared and optical photometric follow-up observations with 8-meter class telescope of this stream, will be 
presented in a companion paper (Martinez-Delgado et al. in preparation).

\subsection{Extracting data and uncertainties}

We use the data in the luminance filter to extract a map of the surface brightness of the 
stream and of the associated uncertainties. Note that we do not need a calibrated map of the absolute luminosity of the stream, but only 
a map of the projected locus occupied by the streaming material, and of the relative brightness of different 
regions along such locus. In order to simplify the comparison with models, we construct 
this map by rasterising the original image, i.e. by dividing the area covered by the data in larger `effective pixels'.
We associate an average brightness -- with uncertainty -- to each of these.
There are two main difficulties to overcome in doing this: 
(i) isolating and excluding all compact bright sources present in the data, such as foreground stars and
background galaxies; (ii) estimating and accounting for the smooth variations in the non-flat background 
luminosity.

We solve issue (i) using a twofold strategy. First we take much care in manually masking out all 
foreground stars and background galaxies brighter than some threshold. The result of this operation is illustrated in
the left panel of Fig.~\ref{masks}, which shows the resulting mask (excluded region in white). This concurrently masks out
the host galaxy, the progenitor of the stream and the region in the upper left corner of the data. The reason
for excluding the latter portion of data is that this contains the poorly determined apocenters of the two 
upper plumes of the stream. Because of scattered light from a bright foreground star and the intrinsically low surface
brightness of this section of the stream, the exact apocenters of these plumes are highly uncertain. 
As a consequence, rather than imposing incorrect constraints through possibly inaccurate data, 
we prefer to leave complete freedom as to the extension of these plumes, and mask this region out. Second, in the original image, 
we systematically exclude all single pixels that are brighter than a given threshold. This is considerably higher than values 
reached within the stream, so this final step simply takes care of masking all spurious sources that had survived the first mask. 
We refer to the final non masked region using the symbol $\mathcal{M}$.

In order to subtract the background, we smooth it through a classical Gaussian kernel. 
For a given size of the effective pixels, we calculate average brightnesses and uncertainties using the following classical estimates
\begin{align}
\nu_i & =\langle\mu \rangle_i-\langle\mu \rangle_{\mathscr{G}, i}\ , \\
N_i\; \delta\nu_i^2 & = (\langle\mu^2 \rangle_i-\langle\mu \rangle_i^2)+(\langle\mu^2 \rangle_{\mathscr{G}, i}-\langle\mu \rangle_{\mathscr{G}, i}^2)\ ,\label{uncert}
\end{align}
where $\nu_i$ indicates the brightness of the effective pixel $i$ and $\delta\nu_i$ its uncertainty, while $\mu$ is the brightness 
of pixels in the original data; $\langle\cdot \rangle_i$ is the average over the area of the effective pixel $i$ and $\langle\cdot \rangle_{\mathscr{G}, i}$
indicates the average centred on the effective pixel $i$ and obtained using a Gaussian kernel -- with characteristic size a 
few times larger than the size of the effective pixel itself. Note that both averages only use pixels in the original image that are not 
masked out, i.e. pixels in $\mathcal{M}$, so that $N_i$ is the number of such pixels used for the effective pixel $i$.

\begin{figure}
\centering
\includegraphics[width=.85\columnwidth]{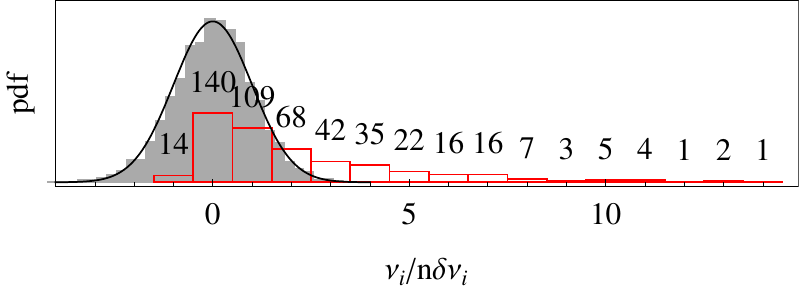}\vspace{.3cm}
\includegraphics[width=.49\columnwidth]{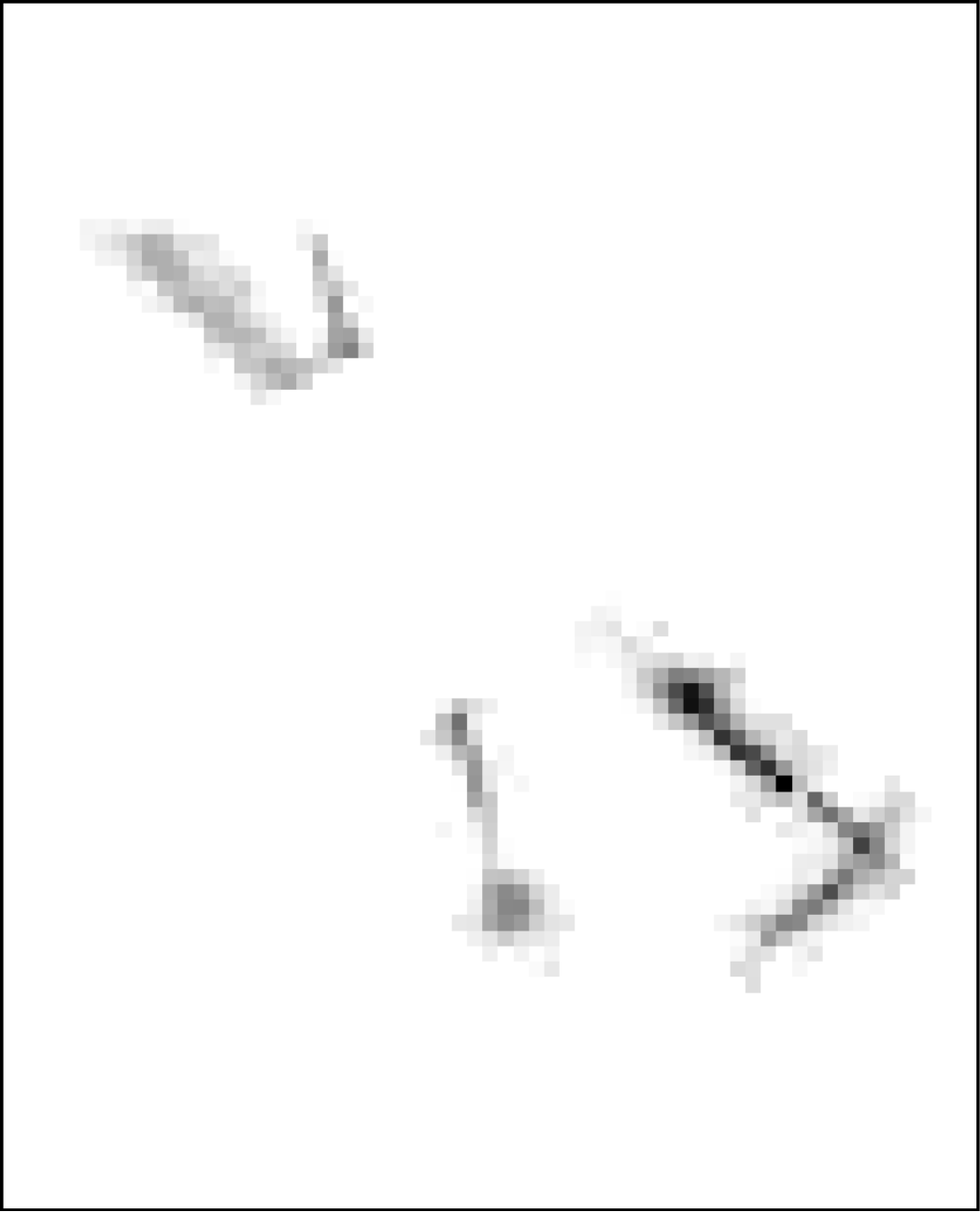}
\includegraphics[width=.49\columnwidth]{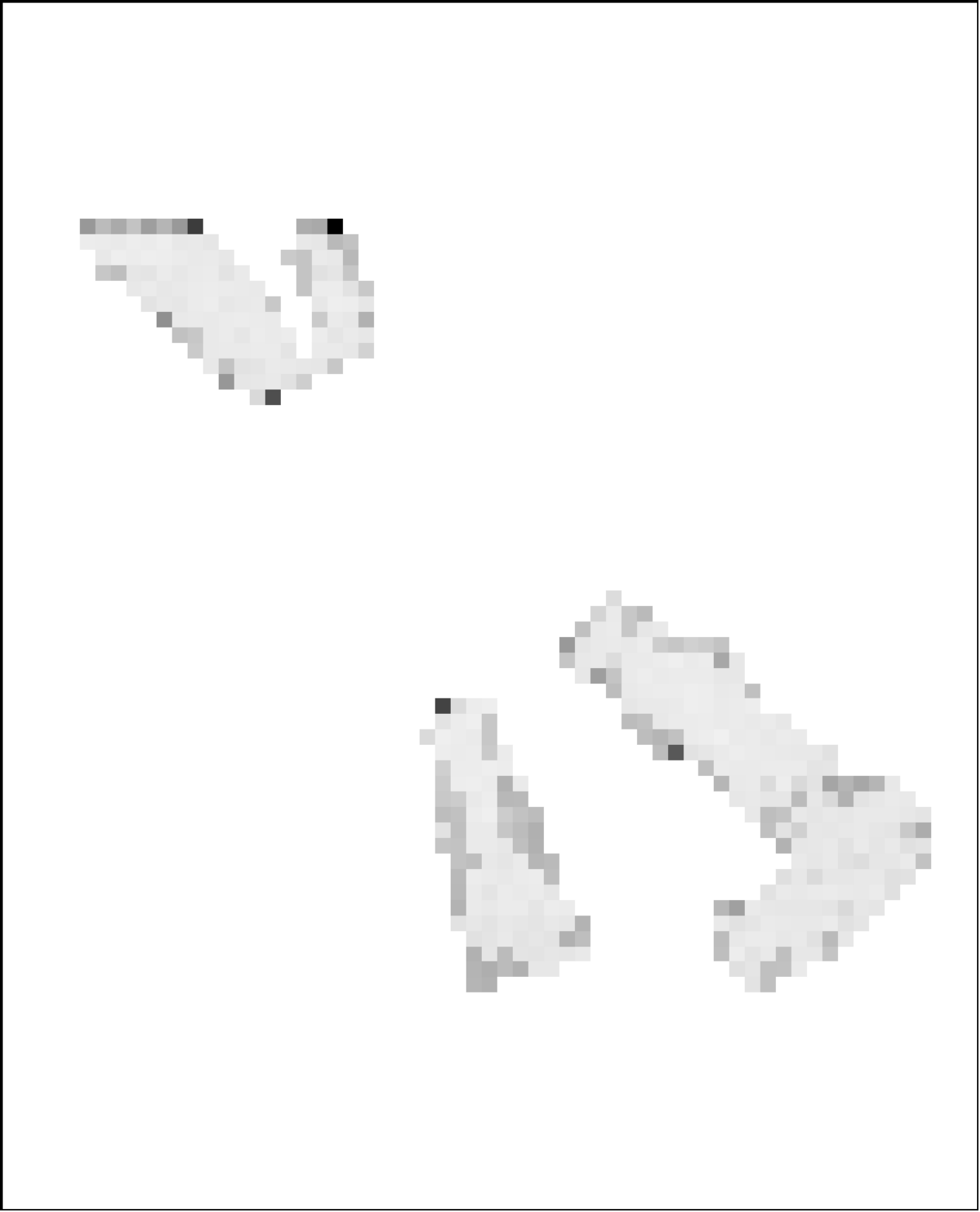}
\caption{{\it Upper panel:} the distribution of the noise in the region where the stream is absent, in grey, together with 
an histogram of the significance of the signal in the region of the stream (see Fig.~\ref{masks}).
{\it Lower panels:} the map of the surface brightness in the region of the stream ({\it left}),
is shown together with the corresponding uncertainties ({\it right}).
\label{data}}
\end{figure}

So to learn about the properties of the noise, we separate the available data in two regions, as from the mask in the right 
panel of Fig.~\ref{masks}, isolating the stream in the region $\mathcal{M}_{\rm str}$.
In the complementary region, masked out in the right panel of Fig.~\ref{masks}, we impose that 
\begin{equation}
{\rm p}\left({{\nu_i}\over{n\: \delta\nu_i}}\right)\approx\mathscr{G}(0,1)\ ,
\label{gnoise}
\end{equation}
where ${\rm p}$ indicates the probability distribution of the brightness of the effective pixels, normalised by their uncertainty.
In other words, we rescale the uncertainties calculated using eqn.~(\ref{uncert}) so that the distribution in eqn.~(\ref{gnoise}) is as similar as possible to 
a standardised Gaussian, as it would be if outside $\mathcal{M}_{\rm str}$, where the stream is absent, the image is dominated by noise. 
In practice we use 
\begin{equation}
n=\left[\left\langle \left({{\nu_i}\over{\delta\nu_i}}\right)^2 \right\rangle-\left\langle {{\nu_i}\over{\delta\nu_i}} \right\rangle^2 \right]^{1\over 2}\ ,
\label{qnoisesc}
\end{equation}
and find that, for an effective pixel size that is a fraction of the stream's width, values $n\approx 3$ are needed.  
The grey shaded area in the upper panel of Fig.~\ref{data} shows the resulting distribution of the `signal' in the region where the stream is absent, masked 
out in the right panel of Fig.~\ref{masks}. This well compares to a Gaussian distribution with zero mean and unit standard deviation,
also shown as a full black line. Therefore, we apply the scaling required by eqn.~(\ref{qnoisesc}) to all effective pixels, including those in the region 
containing the stream, $\mathcal{M}_{\rm str}$.

The final result of this procedure is shown in the lower panels of Fig.~\ref{data}: using the same grayscale for both average brightness and 
associated uncertainties, these illustrate the map of the effective pixels in the stream's region $\mathcal{M}_{\rm str}$. The stream's morphology 
can be clearly recognised, with values $\nu_i/n\delta\nu_i$ reaching up to over 14 for this particular effective pixel size, corresponding to a side of 2 kpc 
in the sky at the distance of NGC 1097 \citep[D=16.96 Mpc, ][]{Ko04}. The effective pixels characterised by significant uncertainties are due to the Poisson noise of eqn.~(\ref{uncert}):
as they are close to regions masked out of $\mathcal{M}$, $N_i$ decreases considerably. The complete distribution in $\nu_i/n\delta\nu_i$
of the region $\mathcal{M}_{\rm str}$ is shown in the upper panel of Fig.~\ref{data} with a (scaled) red histogram.

Finally, we use the original image to measure the current position of the remnant of the progenitor in the plane of the sky. 
With respect to centre of NGC 1097, we find 
\begin{equation}
(x_{{\rm p}, e_1}, x_{{\rm p}, e_2})=(51.5, -33,8)\pm(0.3, 0.3)\ {\rm kpc}\ ,
\label{progproj}
\end{equation}
where the directions $\bm{\hat{e}_1}$  and  $\bm{\hat{e}_2}$ are respectively the horizontal and vertical directions in Fig.~\ref{rawdat}, in the plane of the sky.
This corresponds to a projected distance of
\begin{equation}
R_{\rm p}=61.6\pm0.4\ {\rm kpc}\ . 
\label{progproj2}
\end{equation}
%

\section{Stream Generator}

\subsection{The technique and the free parameters}
We refer to \citet{NA15}, for brevity A15, for a complete description of the method used to generate mock tidal
streams, and report here a brief summary of the strategy and a description of any additional feature.

As in A15, the underlying total gravitational potential is generated by a spherically symmetric broken power law profile
\begin{equation}
\rho(r)=\rho_0 \left({r\over r_0}\right)^{-\gamma_i}\left[1+\left({r\over r_0}\right)^2\right]^{-{{(\gamma_o-\gamma_i)}/ 2}}\ ,
\label{bpwl}
\end{equation}
with inner and outer slopes $\gamma_i$ and $\gamma_o$, and break radius $r_0$. Orbits in such potential are described using 
a purposely optimised orbit library, that covers the parameter space
\begin{equation}
\left\{
\begin{array}{rcl}
\gamma_i&\in& [0,2.9] \\
\gamma_o&\in& [2.1,5] \\
j&\in& [0,1] \\
\log_{10}(r_c / r_0)&\in& [-2,2] 
\end{array}
\right.\ ,
\end{equation}
where $j$ is the classical circularity and the ratio $r_c/r_0$ is used as a convenient substitute
to orbital energy, $r_c(E/G\rho_0 r_0^2)$ being the radius of the circular orbit with the dimensionless 
energy $E/G\rho_0 r_0^2$. As fully described in A15, this technique bypasses all `on the fly' integrations of the differential equations
of the motion and, as a consequence, is such that evolution of a set of particles to any given time is equally expensive.

Mock streams are obtained by releasing particles along the orbit of the progenitor $(r_{c,{\rm p}}/r_0, j_{\rm p})$,
starting from infall (operatively, the first apocenter) and until the present time, according to a parametrized 
shedding history ${\rm p}_{\rm sh}(t)$. This is the sum of Gaussian distributions, one for each pericentric passage, with free
mean shedding time, characteristic spread and shed mass. 

The initial conditions at escape $(\bm{\delta r},\bm{\delta v})$ can also be varied at will, so to mimic the mass evolution of the progenitor
or the disruption of a system with internal angular momentum. Particles are released in the proximity of the
instantaneous tidal radius of the progenitor
\begin{equation}
\rt = \left({{G m}\over{\Omega^2-{\partial}^2 \Phi /{\partial}r^2}}\right)^{1\over3}\ ,
\label{tidrad}
\end{equation}
the size of which depends on the progenitor mass $m$ and on the instantaneous galactocentric distance. Particles leave 
the progenitor from apertures in the surface of the effective potential, the size of which is connected to their energy, or 
equivalently to the escape velocity $\bm{\delta v}$.
Estimates for this link have been provided in A15, and we report them here for reference: 
\begin{equation}
\left\{
\begin{array}{lrc}
\bar\varpi_{\varphi} & = & \rt \sqrt{{\rt |\bm{\delta v}|^2}\over{G m}}\ , \\
\bar\varpi_{z} & = &\bar{\varpi}_{\varphi}   \sqrt{\gamma_i \over{\gamma_i+1}}
\end{array} \right. \ ,
\label{semiax}
\end{equation}
where $\bar\varpi_{\varphi}$ and $\bar\varpi_{z}$ are the semiaxes of the elliptical escape apertures,
lying in the plane perpendicular to the radial direction $\bm{\hat r}$, crossing $\rt$. $\bm{\hat\varphi}$ is the angular 
direction within the orbital plane, while $\bm{\hat{z}}$ is perpendicular to the orbital plane. As in A15, these 
quantities are used as characteristic scales of Gaussian probability distributions, describing the pdfs of the spatial
initial conditions at escape (see eqns.~(60-62) in A15).

As to the kinematic escape condition, particles are lost
with the same instantaneous angular velocity of the progenitor \citep{AK12, SG14}, and an additional velocity component $\bm{\delta v}$.
Its cylindrical components -- referred to the orbital plane -- have a Gaussian distribution with adjustable mean and dispersion. Using the notation of A15, we find that
$\sigma_r=\sigma_\varphi=\sigma_z=\sigma_{\rm s}$ is sufficient for the data we have in hand, so that the spread
in the kick velocities is described by a single function of time, $\sigma_{\rm s}$. On the other hand, we need
flexibility in the means of the different components. In particular, we use the following parametrisation 
\begin{equation}
\left\{
\begin{array}{rcrl}
\langle\delta v_\varphi \rangle& = & \delta V_{\rm rot} \cos(\phi_{\rm incl})  &\\
\langle\delta v_z\rangle & = & \delta V_{\rm rot} \sin(\phi_{\rm incl}) &\cos[\varphi(t)-\phi_{\rm orient}]
\end{array}
\right.\ ,
\label{rot}
\end{equation}
where $\delta V_{\rm rot}$ phenomenologically corresponds to internal rotation in the progenitor and is a function of time; 
$\phi_{\rm incl}$ indicates the inclination between internal and orbital angular momentum, and  
$\phi_{\rm orient}$ allows us to change the initial orientation of the disky progenitor, as illustrated in A15.
$\varphi(t)$ is the instantaneous orbital location of the progenitor.

It is worth recalling that we only model the release of `stellar particles' and our mock streams will be used to fit the
stellar stream alone. Dark matter is also lost together with stars: the dark stream is likely much hotter and phase-mixes 
more rapidly as a result of warmer initial conditions, as the halo is more extended and has an accordingly warmer 
kinematical profile within the progenitor. 

Time dependences are treated in a non parametric way. Time dependent free parameters are:
(i) $m(t)$, which implies the size of the tidal radius;
(ii) $\sigma_{\rm s}(t)$, which is kept independent of the progenitor mass, and fixes the spread in the kick velocities at escape;
(iii) $\delta V_{\rm rot}(t)$, which approximately fixes the time when any disk component of the progenitor is disrupted and
lost, and the magnitude of such signature in the escape velocities.
For $m$ and $\sigma_{\rm s}$ we simply use a smooth, spline interpolation between the values of these functions
at each pericentric passage, which are treated as free parameters. We find that this is enough for our scopes and the current data, 
and we do not need to increase the number of free parameters by using a finer mesh in time. 
For $ \delta V_{\rm rot}$, instead, we possibly need to resolve finer time scales, and therefore use a different approach.
We parametrize $ \delta V_{\rm rot}(t)$ as the sum of two Gaussians in time, with free means, spreads and maximum
rotational velocities. This allows us to phenomenologically capture at the same time the quick disruption of a fragile disk, and/or
the durable effects of a rotating compact progenitor.

In conclusion, the free parameters and free functions of the model are listed below.
\begin{itemize}
\item{The progenitor's orbit $(r_{c,{\rm p}}/r_0, j_{\rm p})$ and the time spent since infall, i.e. the first apocenter, in units of the progenitor's orbital time, $t_{\rm inf}/T_{r,{\rm p}}$.}
\item{The progenitor's line-of-sight position with respect to the centre of NGC 1097 at the current time, $x_{\rm p, los}$ -- the projected position 
of the progenitor in the sky is fixed by the data within its uncertainty, see eqn.~(\ref{progproj}). A possible rotation of the orbital plane,  $\theta_{\rm orb}$,
around the vector identifying the position of the progenitor at the current time; in practice, this fixes the inclination of the orbital plane with respect to the line-of-sight direction.}
\item{The host total density profile, parametrized by the two parameters $\gamma_i$, $\gamma_o$. The characteristic
density $\rho_0$ is not directly a free parameter as data are limited to relative surface brightness values only, and no kinematic information on the stream is available.
At the same time, the break radius $r_0$ is only apparently a free parameter: once the energy of the progenitor's orbit and the three-dimensional position
of the progenitor are chosen, $r_0$ is univocally determined.}
\item{The shedding history ${\rm p}_{\rm sh}(t)$, i.e. three free parameters for each Gaussian pdf per pericentric passage, 
minus one, as the integral of the complete pdf is normalised to unity.}
\item{The mass evolution of the progenitor $m(t)$, i.e. the mass itself -- or equivalently the tidal radius -- at each pericentric passage. Note that 
this is treated independently from the shed luminous mass at the previous point.}
\item{The velocity dispersion at escape $\sigma_{\rm s}(t)$, i.e. its value at each pericentric passage.}
\item{The ordered kick velocities at escape $\delta V_{\rm rot}(t)$, i.e. three free parameters for each of the two used Gaussian functions plus two angles for 
inclination and orientation, $\phi_{\rm incl}$ and $\phi_{\rm orient}$.}
\end{itemize}
\begin{figure}
\centering
\includegraphics[width=.65\columnwidth]{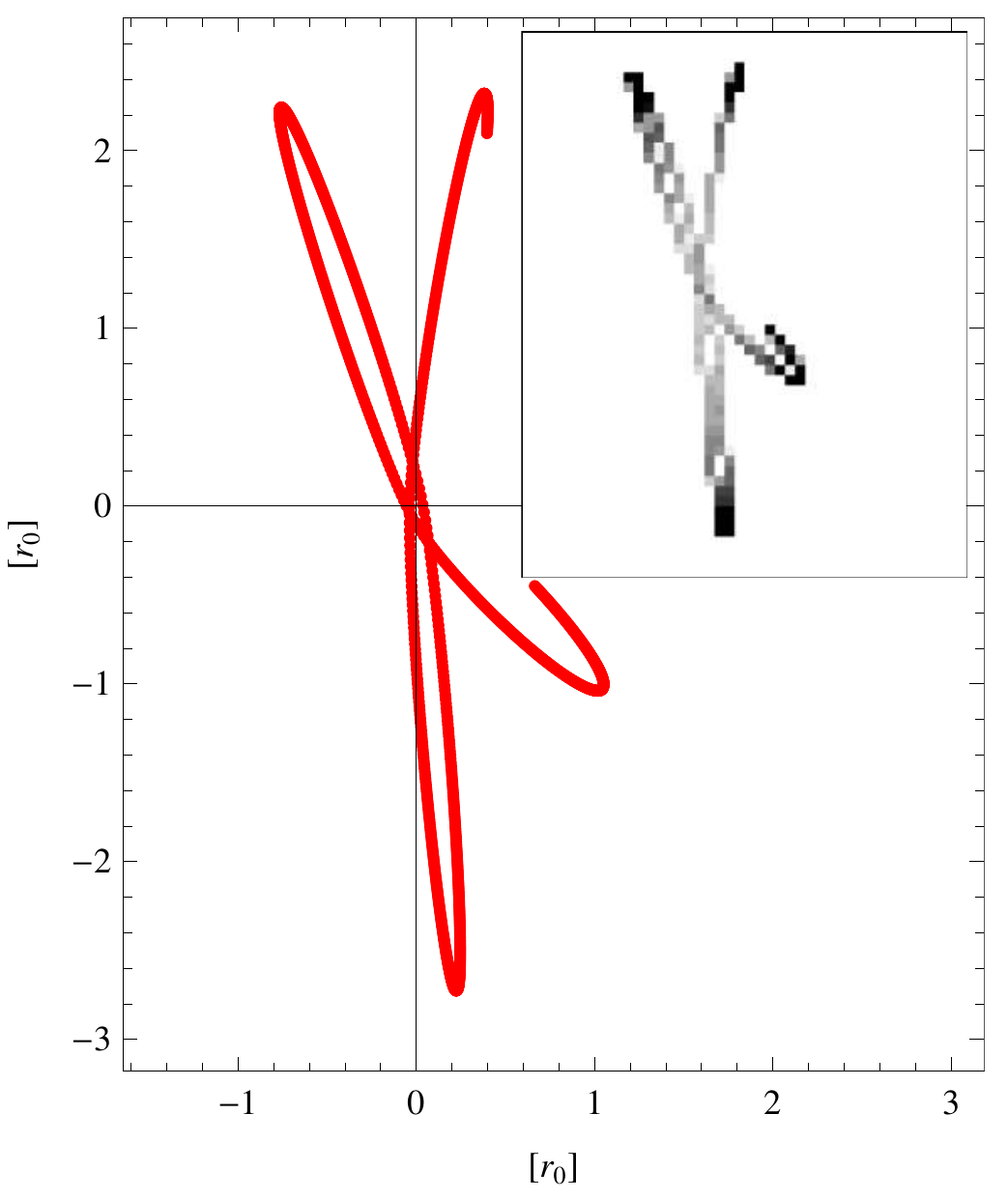}
\caption{A typical dataset used to test the constraining power of the projected shape 
of a single orbit. The large panel illustrates the projected orbit itself, while the inset
shows its corresponding rasterised surface brightness map. \label{sphdata}}
\end{figure}
\begin{figure}
\centering
\includegraphics[width=.85\columnwidth]{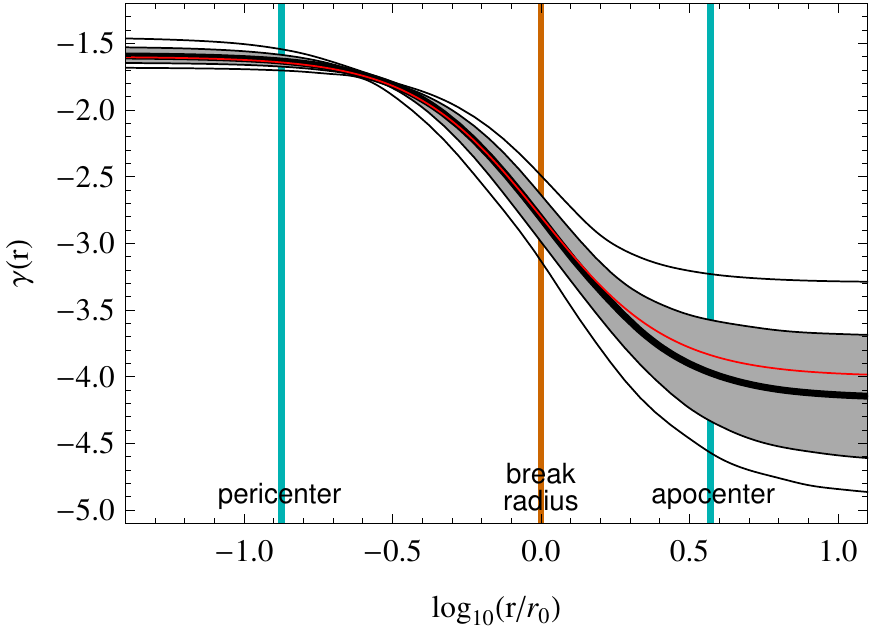}
\caption{The inference on the profile $\gamma(r)$ of the total density, 
recovered from fitting the single orbit displayed in Fig.~\ref{sphdata}.
The grey shaded area represents the 68\% confidence region, the 95\%
confidence region is not shaded. The red profile illustrates the target 
profile, while the vertical lines display the locations of pericenter, break
radius and apocenter.
\label{sphgamma}}
\end{figure}
\begin{figure*}
\centering
\includegraphics[width=1.25\columnwidth]{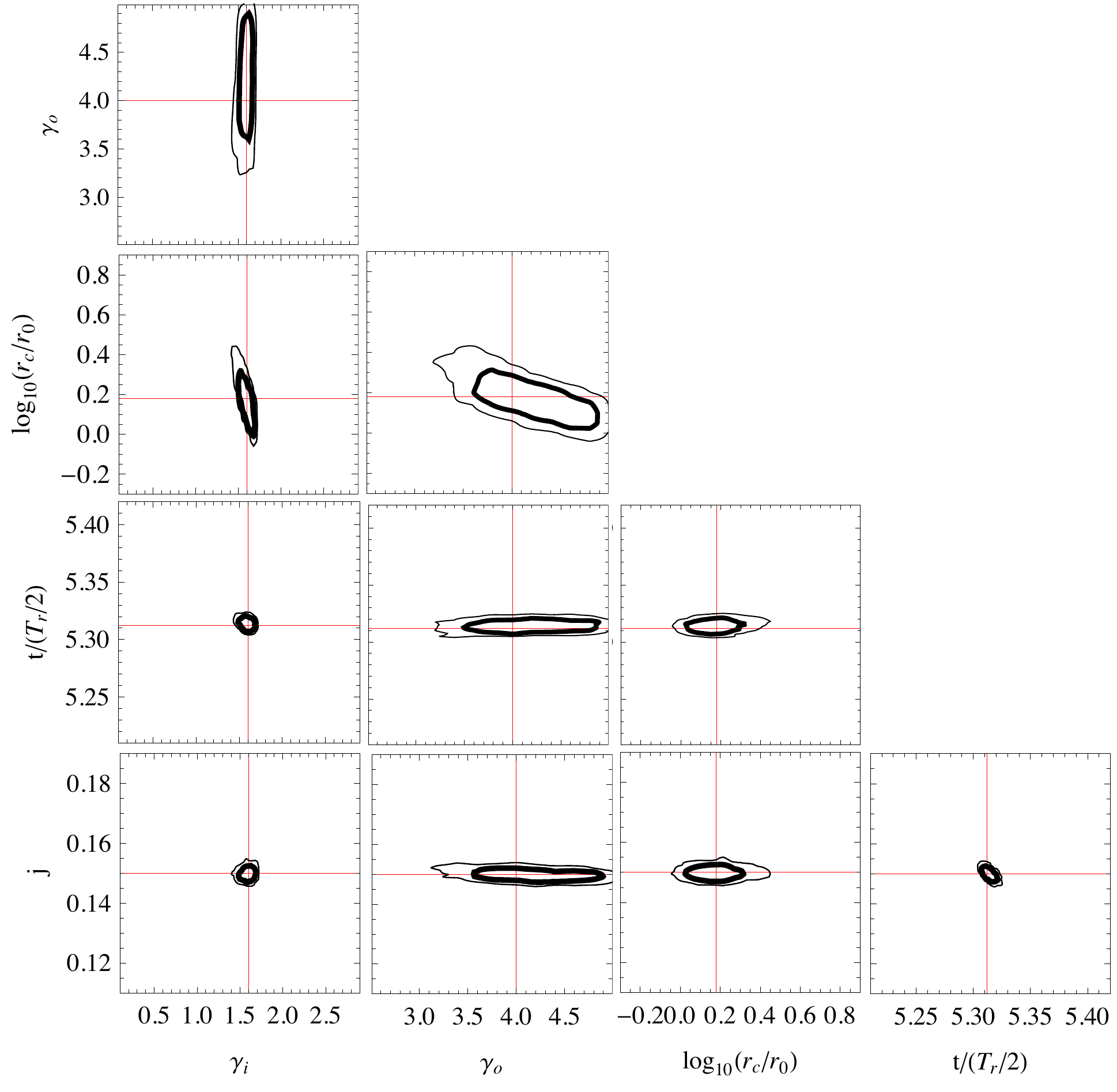}
\caption{Corner plots for the parameters recovered fitting the single orbit displayed in Fig.~\ref{sphdata}.
Contours indicate the 68\% and 95\% confidence regions, while red lines identify the target values. 
\label{sphcorner}}
\end{figure*}
%

\subsection{Likelihood and model comparison}

We populate each stream with $N_{\rm str}=1.25 \times 10^4$ particles, scale it and and rotate it so that the position of the 
progenitor in the plane of the sky at the current time coincides with the one observed in the data. The number of particles in each 
effective pixel $n_i$ is calculated, and transformed into a quantity that can be directly compared to the data.
%
%
This transformation uses an estimate of the surface brightness per particle $\nu_{\rm pp}$,
which is obtained from the region of the stream $\mathcal{M}_{\rm str}$, highlighted in Fig.~\ref{masks}:
\begin{equation}
\nu_{\rm pp} = \sum_{\mathcal{M}_{\rm str}}\nu_i \bigg/ \sum_{\mathcal{M}_{\rm str}}n_i\ .
\end{equation}
For each model $\mathscr{M}$, identified by its set of free parameters, this produces the likelihood
\begin{equation}
\mathscr{L}(\mathscr{M})=\prod_i {{\exp\left[ -{1\over 2} {{(\nu_i-\nu_{\rm pp} n_i)^2}\over{\delta\nu_i^2+\nu_{\rm pp}^2 n_i}} \right]}\over{\sqrt{2\pi (\delta\nu_i^2+\nu_{\rm pp}^2 n_i)}}} \ ,
\label{lik}
\end{equation}
based only on the map of the relative surface brightness of the stream and in which the Poisson noise of each 
effective pixel is taken into account together with the observational uncertainties.

Given the granularity of the mock streams, samplings of the pdf functions that define a given model $\mathscr{M}$
produce slightly different estimates of the likelihood $\mathscr{L}(\mathscr{M})$; in other words,  $\mathscr{L}(\mathscr{M})$ is subject to
variations when calculated repeatedly. 
This variability disrupts the natural exploration of the parameter space, as it causes walkers to get stuck in samplings  
that are fortuitously performing better than the average of that model. As a consequence, such variations must be suppressed, 
and we do this by tempering the likelihood. For a given effective pixel size and number of particles in the 
stream $N_{\rm str}$, we use 
\begin{equation}
\log_{5}\left[\sqrt{ \langle\mathscr{L}(\mathscr{M})^{2/ T_{5}}\rangle-\langle\mathscr{L}(\mathscr{M})^{1/ T_{5}}\rangle^2 }\right]= 1 \ ,
\label{liktemp}
\end{equation}
where averages are calculated on repeated samplings of the same model $\mathscr{M}$, which, operatively, is taken to be 
the best fitting model. In this way, oscillations of the tempered likelihood $\mathscr{L}^{1/T_{5}}$ are limited an order of magnitude at most, 
allowing for a smooth evolution of the walkers. For $N_{\rm str}=12 500$ and an effective pixel size of 4 kpc$^2$, we
find that $T_{5}\approx12$ is needed.

\section{Tests on single orbits}

In order to test whether it is possible to constrain the density profile of the host using the 
projected morphology of the stream alone, we perform tests on single orbits. A sample of orbits 
is chosen with properties comparable to those inferred for the progenitor's orbit, which we illustrate in
Section~5. Such orbits are populated with $\approx10^3$ particles and the resulting surface brightness
in the plane of the sky is mapped in effective pixels. In order to mimic the characteristic width of the plumes 
of dog leg stream, the size of the effective pixels used for these tests is taken to be approximately four times 
larger in area than those used to generate the maps of Fig.~\ref{data}. As shown in Fig.~\ref{sphdata}, 
this considerably broadens the projected profile of the orbits.

We fit such mock datasets using single orbits from our library and find that target parameters are always well
recovered. Figure~\ref{sphgamma} illustrates how the measured radial profile of the density slope compares with 
the generating one (in red). The grey shaded area corresponds to the 68\% confidence region, contained between 
the 16\% and 84\% quantiles; the 95\% confidence region is also shown. Vertical lines illustrate the position of 
orbital pericenter and apocenter, together with the break radius of the underlying density profile.

Figure~\ref{sphcorner} illustrates in more detail the 68\% and 95\% confidence regions of the parameters that
describe the main projected morphology of the orbit, and how these compare with the target
values: as shown by the red lines, all values are recovered within one sigma.  Most parameters are not degenerate
and the only noticeable degeneracies are those associated with the profile of $\gamma(r)$. The inner and outer slopes,
$\gamma_i$ and $\gamma_o$, can be slightly varied around the best fitting values together with the energy of the orbit 
$r_c/r_0$, which, as mentioned earlier, is a measure of the the break radius. 
Such degeneracy manifests itself in a clear pinch point in the recovered profile $\gamma(r)$, in Fig.~\ref{sphgamma},
located where the profile first begins to steepen, between pericenter and break radius.

%
%

%
\begin{figure*}
\centering
\includegraphics[width=1.25\columnwidth]{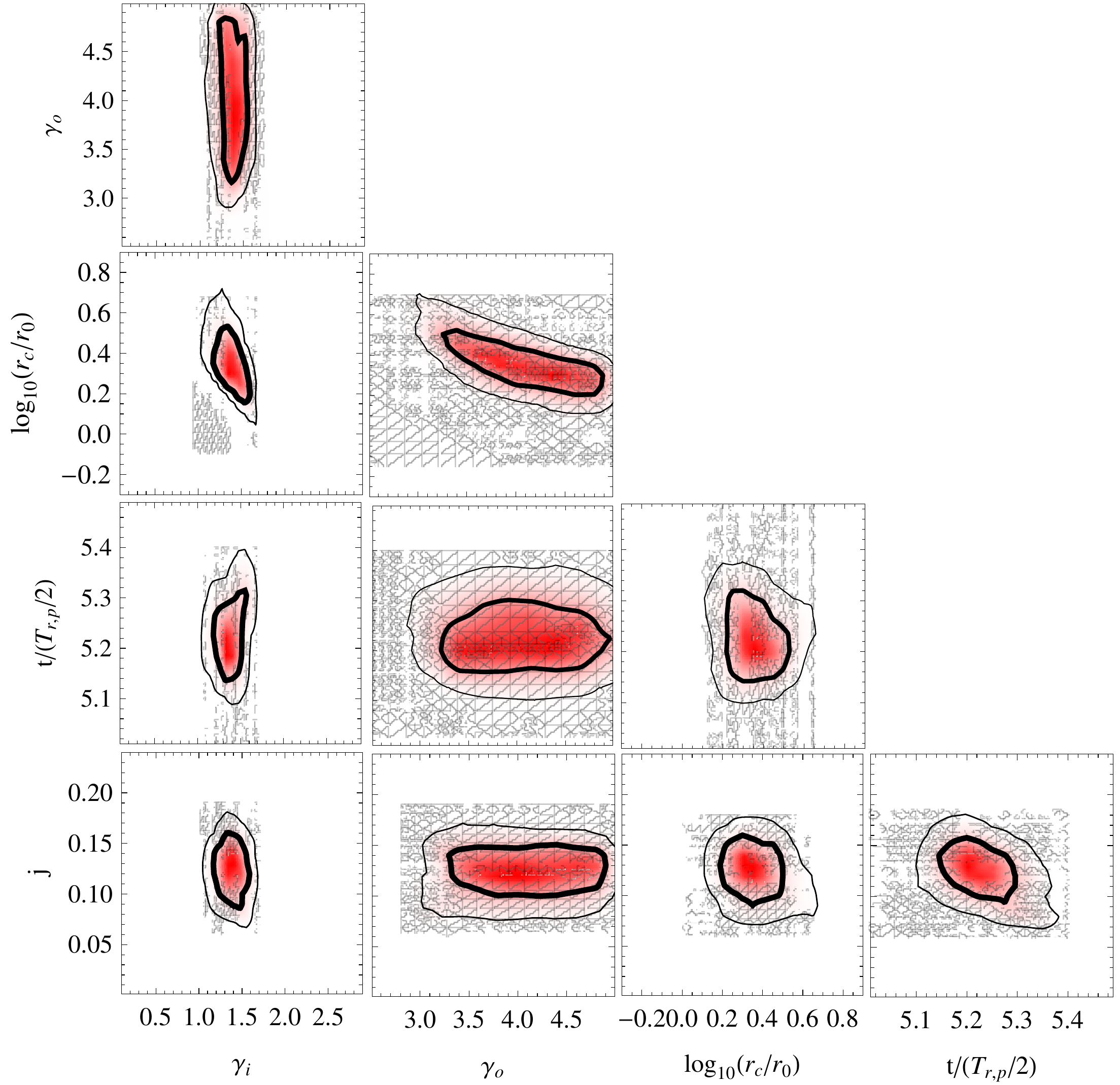}
\caption{Corner plots for the parameters recovered fitting the dog leg stream in NGC 1097: contours indicate the 68\% and 95\% confidence regions. 
\label{fincorn}}
\end{figure*}
%

\section{Results: The host galaxy}

We explore the available parameter space using a suite of Metropolis-Hastings walkers. 
We first perform a series of tentative explorations employing likelihoods temperated with high
temperatures, using multiples of $T_{5}$, defined in eqn.~(\ref{liktemp}). These walkers 
quickly swipe through very wide areas of the parameter space
and approximately identify the useful region. Final runs are performed using $T=T_{5}$.

We repeat the complete analysis on data obtained using two different effective pixel 
sizes -- 4 kpc$^2$ and 3 kpc$^2$ --  so to test the influence of these on any results together 
with the influence of a shifting binning pattern on the image. 
Furthermore, we perform the final runs using two different numbers of particles per stream, 
$N_{\rm str}=1.25\times10^4$ and $N_{\rm str}=7.5\times10^3$. We find that the inferred 
confidence intervals for all parameters are comparable in all cases. There's perhaps a 
tendency for the confidence intervals obtained using a lower $N_{\rm str}$ to be $\approx$10\% wider, 
as a natural consequence of the comparatively higher values of $T_{5}$, which are a result of 
a slightly higher Poisson noise in the effective pixels.

\subsection{The logarithmic total density slope}

Figure~\ref{fincorn} shows the 68\% and 95\% confidence contours of the parameters
that define the qualitative global morphology of the stream. As in the test runs, Fig.~\ref{sphcorner},
the only noticeable degeneracy we identify is between the outer density slope $\gamma_o$ and
the break radius $r_0$, through the orbital energy of the potential $r_{c,{\rm p}}/r_0$. The
asymptotic inner density slope is found to be 
\begin{equation}
\gamma_i=1.4\pm0.15\ ,
\label{gammai}
\end{equation}
where we are reporting median, 16\% and 84\% quantiles, identifying the 68\% confidence interval.
The stream's morphology allows us to clearly see the density profile steepening from this central value, 
and the constraint on the asymptotic value of the outer slope is
\begin{equation}
\gamma_o=4.1\pm0.6\ .
\label{gammao}
\end{equation}
The break radius is at 
\begin{equation}
r_0=37^{+11}_{-9}\ {\rm kpc}\ ,
\label{breakr}
\end{equation}
well within the apocenter of the progenitor's orbit and the total radial extension of the stream.

\begin{figure}
\centering
\includegraphics[width=.95\columnwidth]{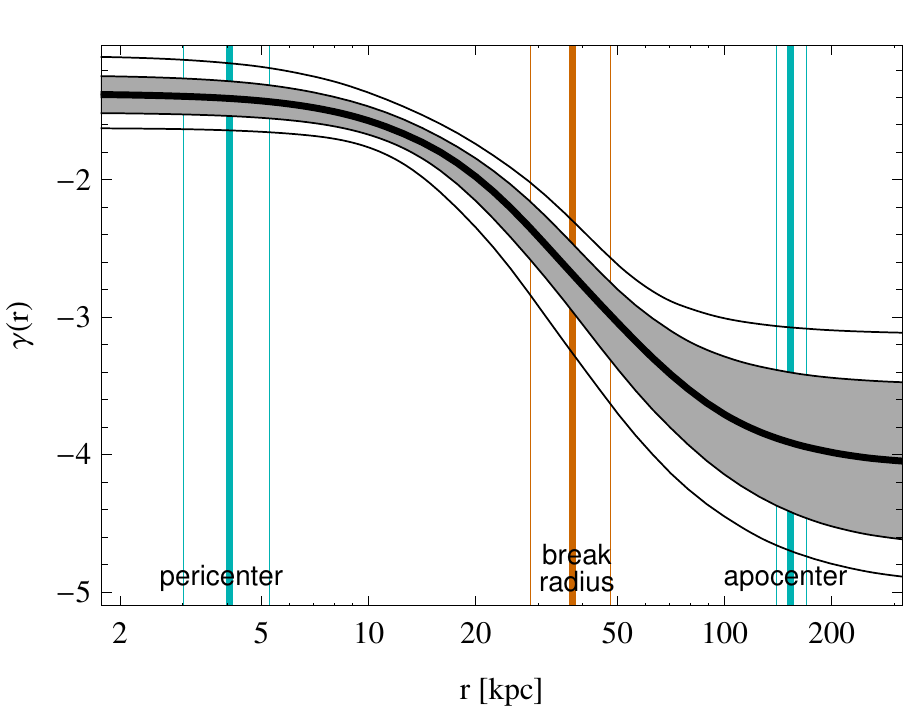}
\caption{The inference on the logarithmic density slope of the total density profile of NGC 1097. 
Grey shading identifies the 68\% confidence region, shown together with the 95\% confidence area.
Vertical lines illustrate how the plotted radial interval compares to the inferred values of orbital pericenter,
apocenter, and break radius of the density profile. For each, the 68\% confidence region is displayed.
\label{gammapr}}
\end{figure}

The full radial profile we infer for the logarithmic total density slope is displayed in Fig.~\ref{gammapr},
where both 68\% and 95\% confidence contours are shown. Vertical lines identify the inferred
pericenter of the progenitor 
\begin{equation}
r_{\rm peri}= 4.1^{+1.2}_{-1}\; {\rm kpc}\ ,
\label{rperi}
\end{equation}
the break radius, and the orbital apocenter
\begin{equation}
r_{\rm apo}= 153^{+17}_{-13}\; {\rm kpc}\ .
\label{rapo}
\end{equation}
The progenitor lies on a significantly eccentric orbit
\begin{equation}
j= 0.12\pm0.03 \ ,
\label{jcirc}
\end{equation}
and has just recently passed apocenter,
\begin{equation}
t/(T_{r,{\rm p}}/2)= 5.23\pm0.07 \ ,
\label{tnow}
\end{equation}
where $t=0$ here identifies the first pericentric passage and $t/(T_{r,{\rm p}}/2)=-1$
is infall, or the first apocenter. It follows that, in the mean time, the progenitor has experienced 
three pericentric passages, at $t/(T_{r,{\rm p}}/2)\in\{0,2,4\}$

\begin{figure*}
\centering
\includegraphics[width=.9\textwidth]{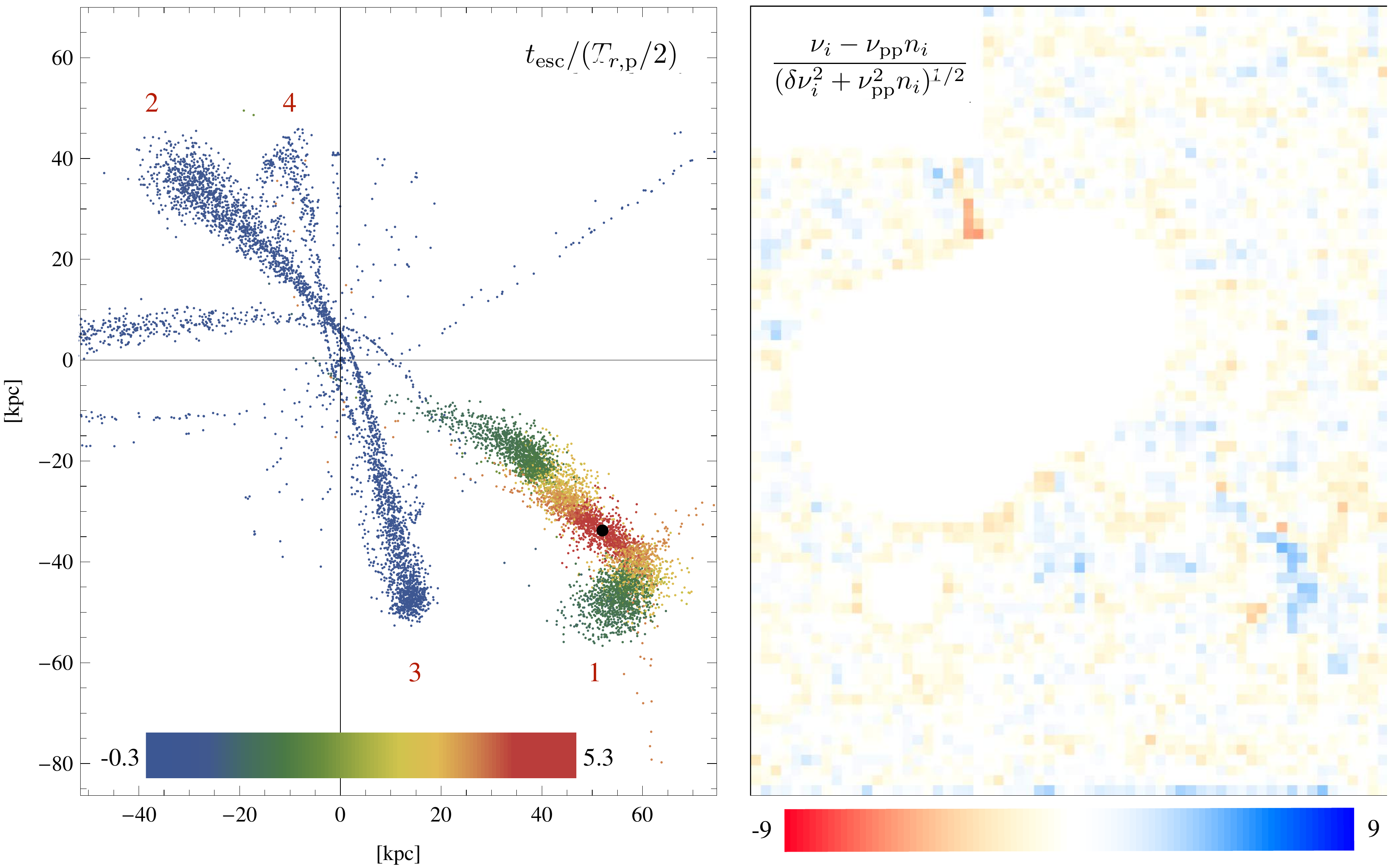}
\caption{{\it Left:} a sampling of the best-fitting model, populated with $N_{\rm str}=1.25\times 10^4$ stream members and 
displayed within the projected region probed by the available data. The current position of the remnant of the progenitor 
is displayed in black. {\it Right} a map of the residuals of the best fitting model, colour coded according to the legend on the right.
\label{bestfitt}}
\end{figure*}

A sampling of the best-fitting model $\mathscr M_{\rm bf}$ is displayed in Fig.~\ref{bestfitt}. The left panel shows the 
position of the progenitor, in black, together with the distribution of stream members, within the
same projected window probed by Fig.~\ref{rawdat}. The stream's morphology is easily recognised,
with its characteristic X-shape and bright `dog leg'. Color coding indicates shedding time as from the 
displayed legend. For ease of reference, we have associated a number to each of the four 
plumes of the stream. {\it Plume 1} is the `leg', contains the remnant of the progenitor, and is a combination of
both leading a trailing material. It collects essentially all stars shed after the first pericentric passage. 
{\it Plumes 2, 3} and {\it 4} are all made of leading material lost at the first pericentric passage, and {\it plume 4} is the farther away 
from the progenitor in terms of differential streaming, and -- as a direct consequence -- also the less extended. 
The right panel is a full map of the model residuals, normalised by the uncertainty of each effective pixel. As illustrated by the colour coding,
residuals are everywhere quite low, rarely above 3 in absolute value, and the only noticeable deficiencies of the model are: (i) a somewhat
reduced brightness in {\it plume 4} (visible in red in the residuals map), and (ii) a mildly puffier  
90 degree turn in {\it plume 1} (evident as a blue clump). We return on this in Section~6.

\begin{figure}
\centering
\includegraphics[width=.45\columnwidth]{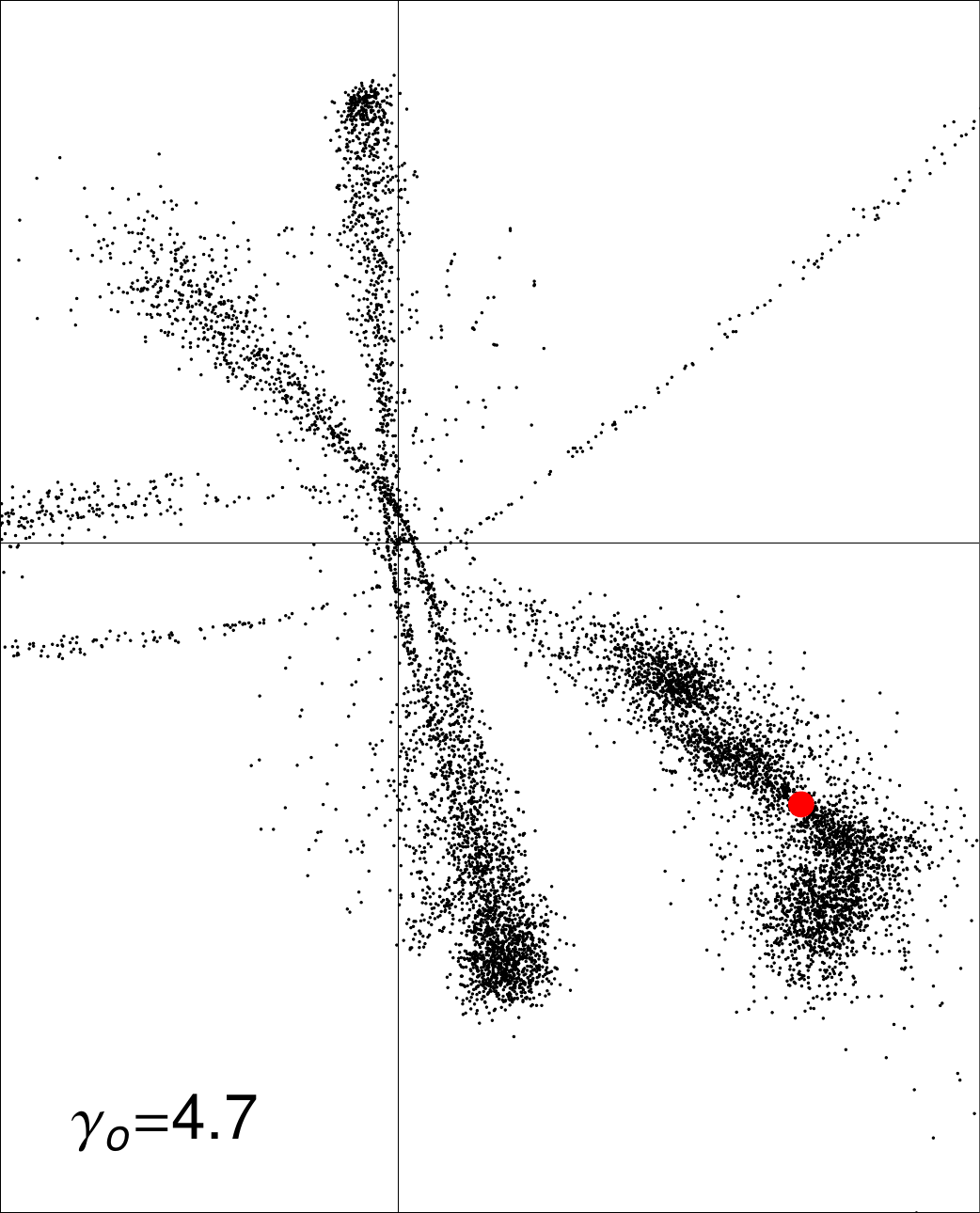}
\includegraphics[width=.45\columnwidth]{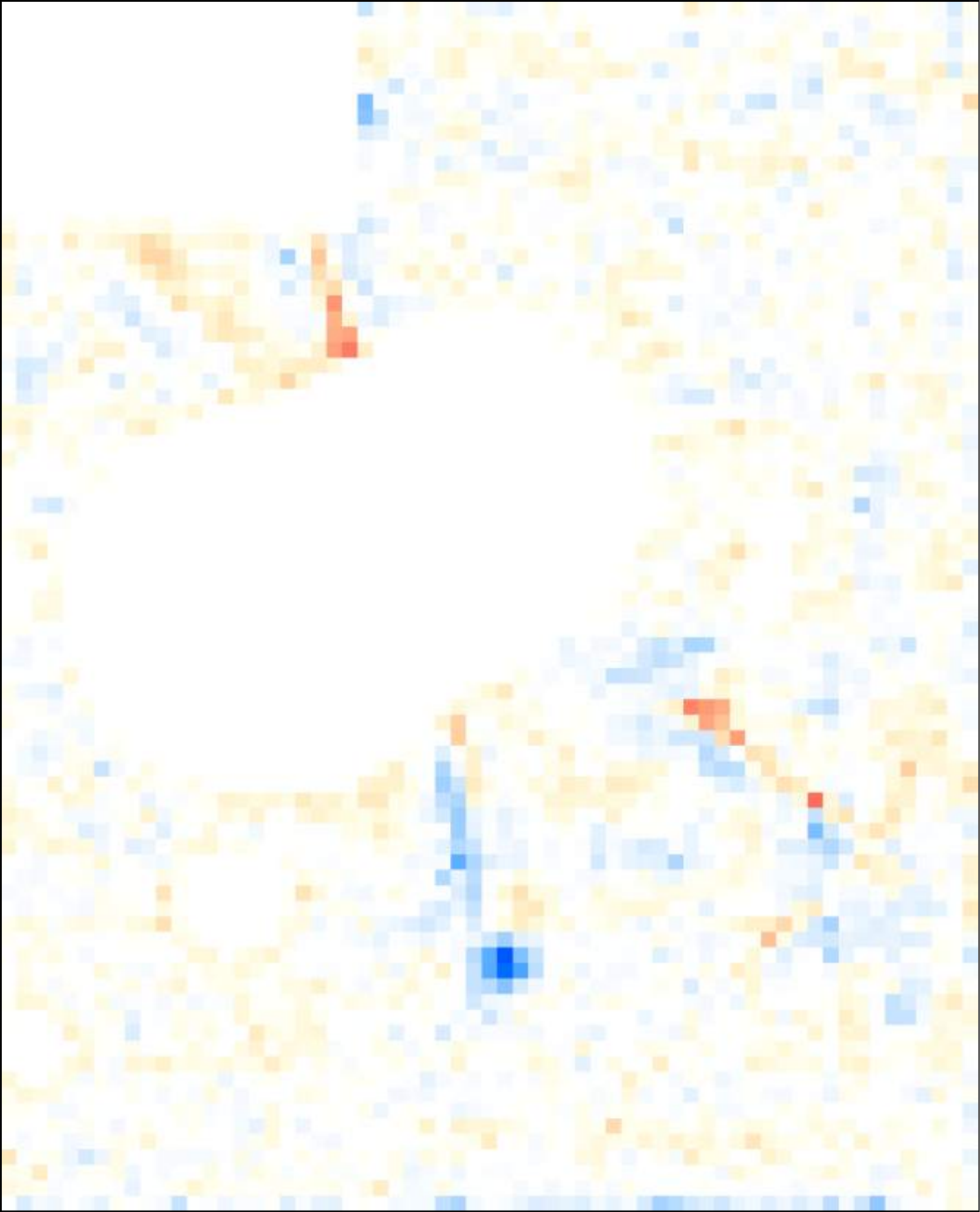}
\includegraphics[width=.45\columnwidth]{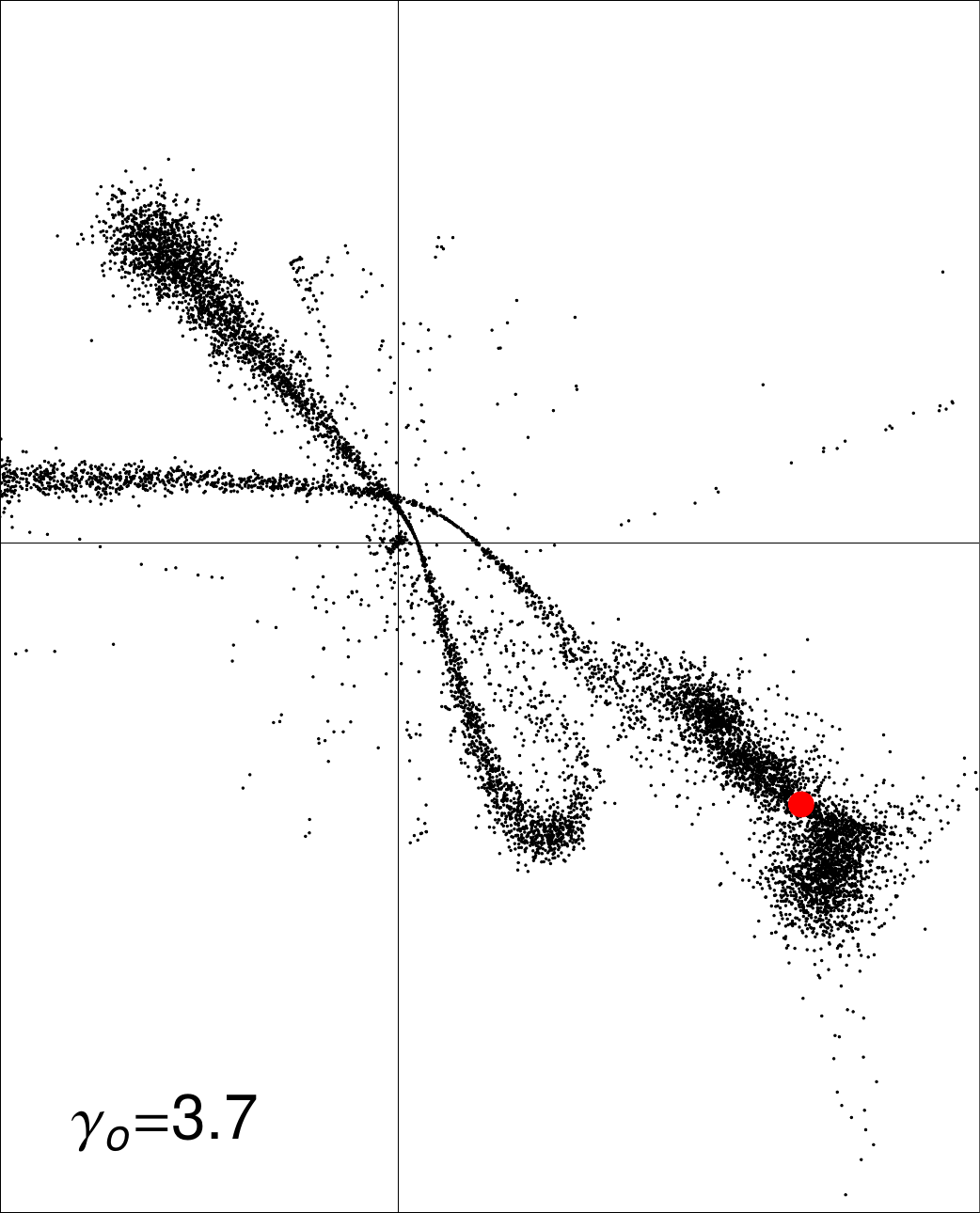}
\includegraphics[width=.45\columnwidth]{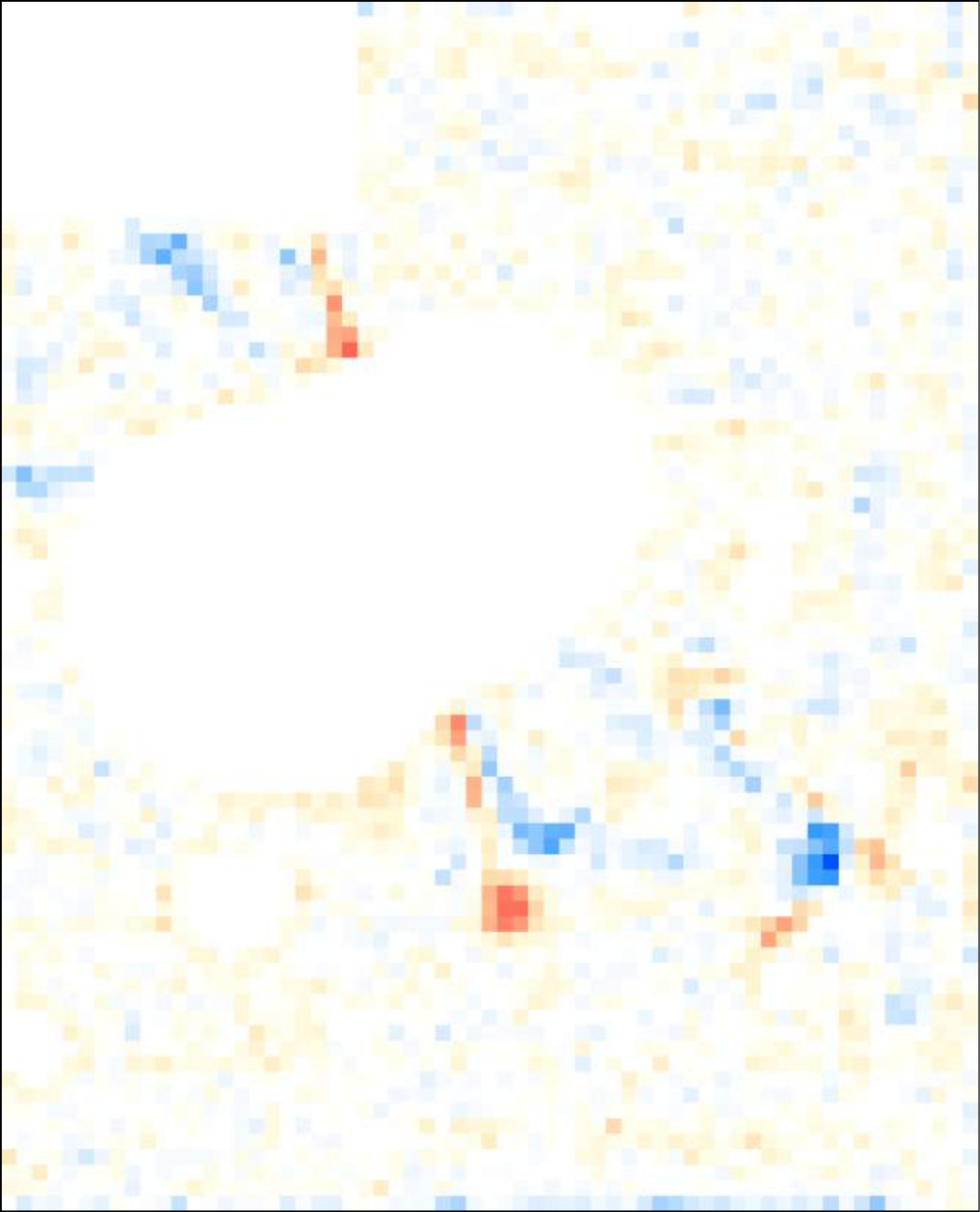}
\caption{An illustration of the qualitative effects on the global morphology of the stream
associated to variations in the outer density slope $\gamma_o$. All other parameters are fixed 
according to the best fitting model (Fig.~\ref{bestfitt}). {\it Upper panels:} the outer slope is steepened to 
$\gamma_o=4.7$, a corresponding stream and its residuals are shown. {\it Lower panels:} the outer slope is
decreased to $\gamma_o=3.7$.
\label{bestfitgo}}
\end{figure}

In order to illustrate in more detail how the stream's projected morphology is capable of constraining the
slope of the total density profile, we show the effect of varing the outer density slope $\gamma_o$, while all other
parameters are fixed as from the best-fitting model $\mathscr M_{\rm bf}$. We explore approximately the 
edges of the 68\% confidence region of $\gamma_o$. As shown in the upper panes of Fig.~\ref{bestfitgo}, 
increasing the outer density slope to $\gamma_o=4.7$: (i) increases the speed of the 
differential streaming, so that {\it plume 4} appears considerably brighter at the expenses of {\it plume 2}; (ii)
implies significant increase in the galactocentric distances of the apocenters 
of the stream's plumes, as for similar orbital energies larger distances can be reached thanks to a 
more slowly rising gravitational potential; (iii) increases the angles between {\it plume 1} and each of the others, as 
a consequence of the change in the shape of orbits. 
The result is that the location and shape of all four plumes is changed enough to result in substantial residuals, 
shown in the upper residuals map with the same colour coding of Fig.~\ref{bestfitt}, much higher than in the case of the best fitting model.

The effect of a lower $\gamma_o=3.7$ is analogous, although changes have the opposite qualitative direction,
as proved by a comparison between the two residuals maps: the streaming speed is reduced and so are the length of the plumes  
and the angles between them.
From this overview we see that the most important single pieces of information
providing the constraints on the outer host density profile are: the length of {\it plume 3} with respect to the current position
of the remnant (or alternatively, to the length of plume 1), and the angles between {\it plume 1} and all the others --
it is worth recalling that the lengths of {\it plumes 2} and {\it 4} are not constrained by the data (see Section~2).

%

\subsection{The total mass profile}

The logarithmic density profile is a dimensionless quantity and can be derived from the surface brightness map of the stream alone. 
To associate a physical scale to the break radius of the profile, a dimensional scale of length is necessary and this is
provided by the projected distance of the progenitor from the centre of NGC 1097,
eqns.~(\ref{progproj}) and~(\ref{progproj2}). However, in order to proceed further, we need a dimensional scale to transform our inference on $\gamma(r)$
in an inference on mass. In order to do so, we use the HI rotation curve measured by \citet{HW03}: we make reference to 
the outer parts of the available rotation curve, which is observed to be approximately flat:
\begin{equation}
V_c(29.6\pm3\ {\rm kpc})=270\pm 10\ {\rm kms}^{-1} \ .
\label{vcbar}
\end{equation}
We use this condition to bring each of the dimensionless models accepted by our walkers to dimensional ones.  
Note that, in all inferences that require use of~(\ref{vcbar}), reported uncertainties are the combination of the uncertainties
deriving from the stream model, together with those propagated by the observational errors of the constraint~(\ref{vcbar}), which 
we assume have a Gaussian distribution.

\begin{figure}
\centering
\includegraphics[width=.49\columnwidth]{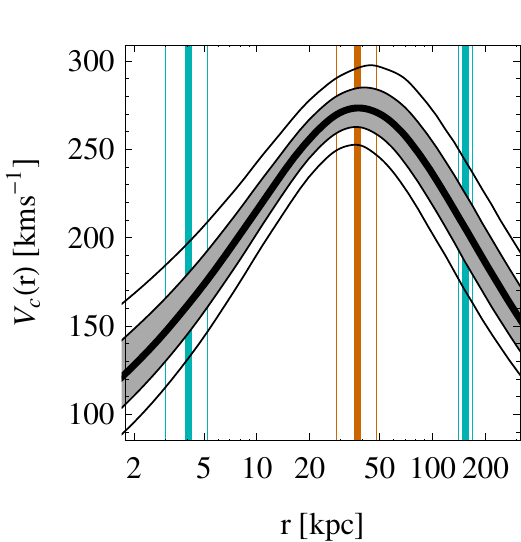}\includegraphics[width=.49\columnwidth]{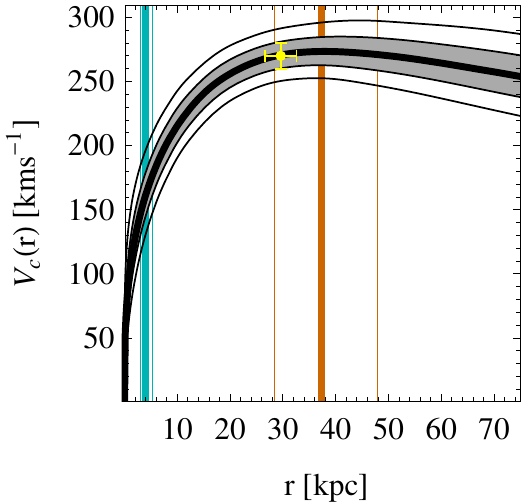}
\caption{The circular velocity profile inferred for NGC 1097 in logarithmic scale ({\it left})
and linear scale ({\it right}). The yellow datapoint in the right panel shows the 
observational constraint eqn~(\ref{vcbar}), used to normalise our dimensionless stream models.
\label{velprof}}
\end{figure}

Figure~\ref{velprof} shows the resulting circular velocity profile of NGC 1097, in both linear and logarithmic scale. 
The yellow point in the upper panel displays the normalisation imposed by~(\ref{vcbar}). As measured by \citet{HW03}, the rotation curve
is approximately flat at those radii. The stream model naturally reproduces this finding and suggests the circular velocity remains flat 
out to $r\gtrsim r_0$. At even larger radii, however, the density profile steepens and the circular velocity 
declines considerably, as illustrated by the left panel.

\begin{figure}
\centering
\includegraphics[width=.9\columnwidth]{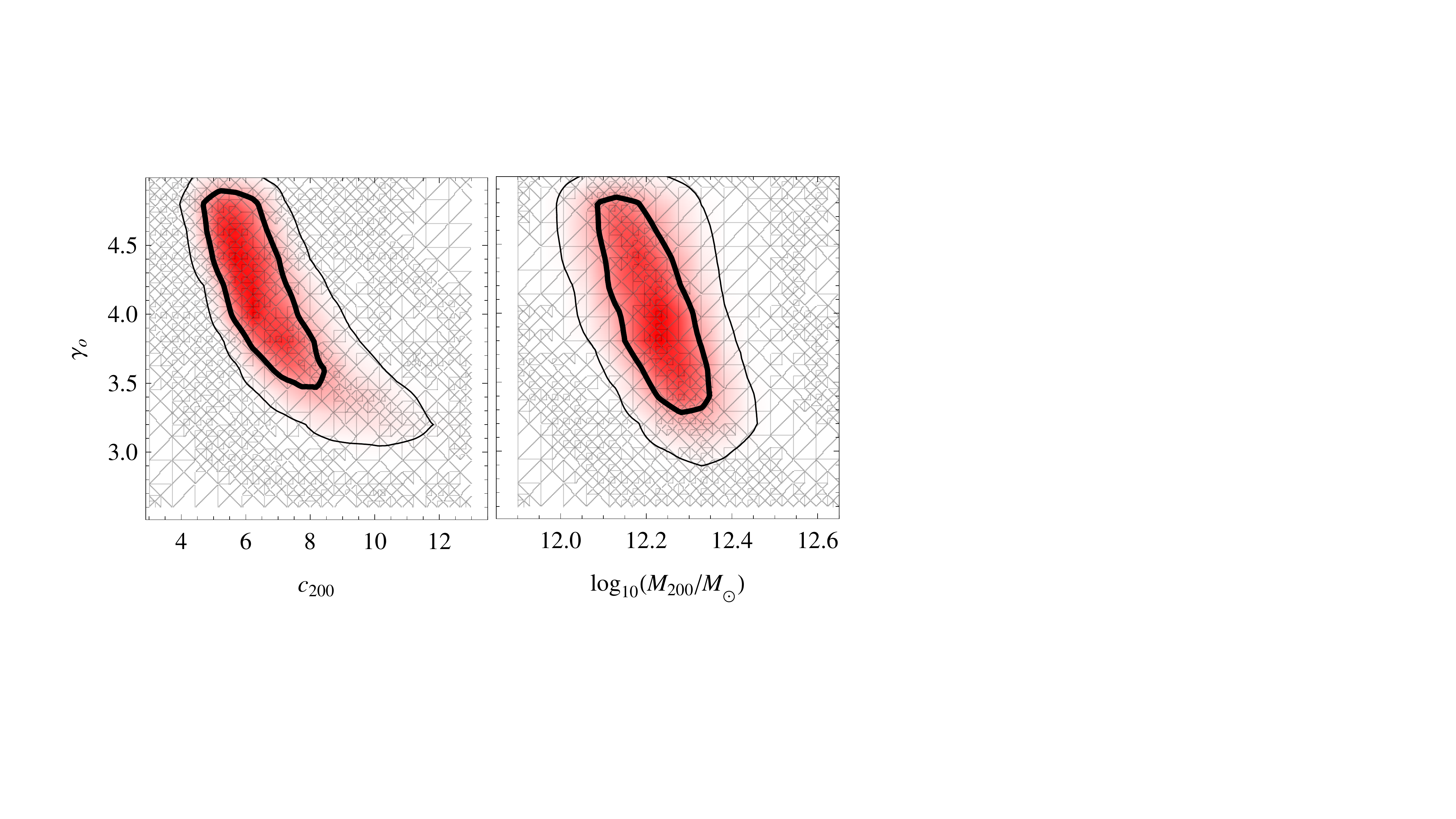}
\caption{The degeneracy between the outer density slope $\gamma_o$ and concentration $c_{200}$, or virial mass $M_{200}$.
68\% and 95\% confidence regions are displayed. 
\label{1dpdfs}}
\end{figure}

The combination of the constraint~(\ref{vcbar}) and our stream model also provides a total mass: NGC 1097 is found to have a sizeable virial mass of 
\begin{equation}
{\rm log}_{10}(M_{200}/M_\odot)=12.23\pm0.12 \ ,
\label{m200}
\end{equation}
and a virial radius of
\begin{equation}
r_{200} = 250^{+21}_{-19}\ {\rm kpc} \ .
\label{r200}
\end{equation}
We do not subtract stellar mass explicitly, and infer the concentration of this halo using the total mass profile. 
This implies 
\begin{equation}
c_{200} = 6.7^{+2.4}_{-1.3} \ .
\label{c200}
\end{equation}
Figure~\ref{1dpdfs} shows the degeneracy that exists between asymptotic outer density slope $\gamma_o$ and concentration, and
correspondingly, virial mass. Steeper density profiles require larger break radii, implying lower values of the concentration. In turn,
less steep density profile can accommodate larger values of the virial mass.

\section{Results: The progenitor}

\subsection{Mass evolution}
The main ingredients in the global morphology of the stream are the density profile of the host, 
which we have described in Section~5, and the mass of the progenitor. By themselves, the density profile and the progenitor's 
orbit determine the approximate shape of the track of the stream. However, the actual
length and width of the streaming material along this track are determined in first instance by the progenitor's mass.
Through the tidal radius~(\ref{tidrad}), this fixes the speed of the differential streaming (see A15),
and therefore the extension of the tidal tails with respect to the progenitor. 

As mentioned in Section~3, we model the time dependence of the progenitor's mass in an essentially non parametric way, 
i.e. by using the masses at each pericentric passage as free parameters, and by interpolating with smooth splines between them.
The progenitor is found to have recently passed the orbital apocenter (see eqn.~(\ref{tnow})), and has experienced three pericentric passages since 
infall. Using eqn.~(\ref{vcbar}), we find that the apocenter preceding the first pericenter happened at
\begin{equation}
t_{\rm inf} = - 5.4=\pm0.6\ {\rm Gyr} \ ,
\label{tinf}
\end{equation}
and we can use this as an estimate of the infall time. 

\begin{figure}
\centering
\includegraphics[width=.9\columnwidth]{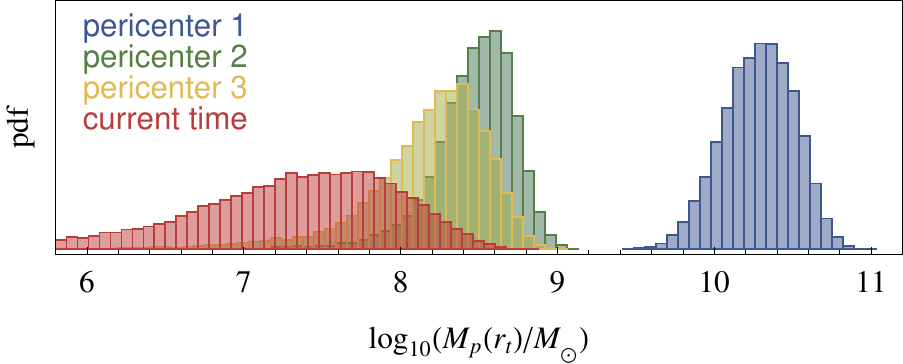}
\includegraphics[width=.9\columnwidth]{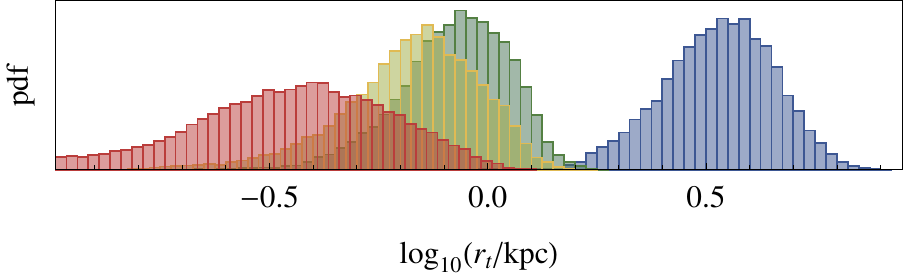}\\\hspace{-.24in}
\includegraphics[width=.95\columnwidth]{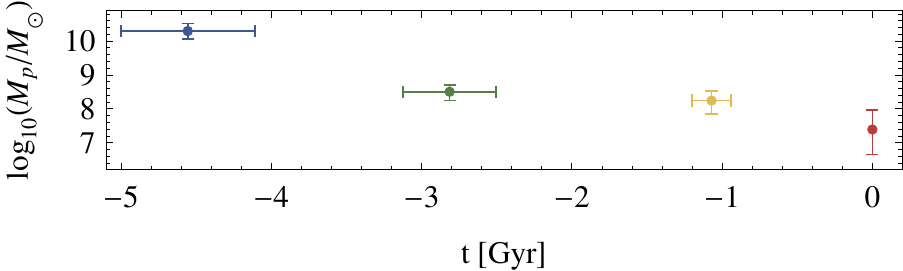}
\caption{The mass evolution of the progenitor of the stream. {\it Upper panel}: 
marginalised pdfs for the total masses of the progenitor within the tidal radius at the 
first ({\it blue}), second ({\it green}), and third ({\it yellow}) pericentric passages, 
together with the mass inferred for the remnant at the current time ({\it red}).
{\it Intermediate panel:} with the same colour coding, the marginalised pdfs for the 
corresponding tidal radii. 
{\it Lower panel}: the total mass evolution in physical time. 
\label{massevol}}
\end{figure}

Since that time, the total mass of the progenitor has evolved considerably:
Figure~\ref{massevol} shows the complete marginalised inference we obtain for the mass evolution of the progenitor. The upper 
panel displays the probability distributions of the total masses at each pericentric passage. For construction, these are
the masses contained within the tidal radius at that time, the probably distribution of which is shown in the intermediate panel. 
At its first interaction with the host, the progenitor is found to have a total mass of
\begin{equation}
\log_{10}\left[m_{{\rm peri}1}(3.4^{+1.1}_{-0.9}\ {\rm kpc} ) /M_\odot\right]= 10.35\pm0.25 \ .
\label{mperi1}
\end{equation}
Such a high mass is required in order to ensure the significant differential streaming of the material shed around the first pericenter.
As illustrated by the left panel of Fig.~\ref{bestfitt}, such material entirely makes up plumes 2, 3 and 4. 

At the second pericentric passage, we find that the total mass is reduced 
by more than 1.5 orders of magnitude: 
\begin{equation}
\log_{10}\left[m_{{\rm peri}2}(0.86^{+0.26}_{-0.23}\ {\rm kpc} ) /M_\odot\right]= 8.5^{+0.2}_{-0.3} \ .
\label{mperi2}
\end{equation}
We interpret this as due to the eccentric progenitor's orbit: tidal forces likely strip
very efficiently the biggest part of the progenitor's dark matter halo when this plunges deep into the 
central regions of its host for the first time.
Luminous material lost around the second pericenter is in the leg of the stream, as shown in Fig.~\ref{bestfitt},
and actually makes up both the peculiar `foot' and puffy `knee'. 
After the first pericentric passage, the surviving remnant is compact enough to be quite resilient to tides, and mass evolution slows down considerably,
with the total mass at the third and last pericentric passage being just slightly lower than at the second (see Fig.~\ref{massevol}).
%
%
Material lost since the second pericenter has a much lower differential streaming and has not gained much distance from the 
progenitor, making up for the straight part of the {\it plume 1}, or the `leg' itself.

Simplifying, these constraints are provided by the distances between progenitor and material shed at each pericentric passage,
together with the extent of such material along the track of the stream. Both quantities are directly related to the speed of the differential streaming,
which is a function of the tidal radius at the time of escape, and then a function of progenitor mass.
It follows that uncertainties on the total mass of the remnant at the current time are comparatively larger, 
as the amount of stream members shed between the last apocenter and the current time is comparatively much smaller:
\begin{equation}
\log_{10}\left[m_{{\rm now}}(0.36^{+0.24}_{-0.16}\ {\rm kpc} ) /M_\odot\right]= 7.4^{+0.6}_{-0.8} \ .
\label{mnow}
\end{equation}
The complete constraints on the mass evolution of the progenitor are collected in the bottom panel of
Fig.~\ref{massevol}, which shows the total mass $m$ at each of the
three pericentric passages, and today. Here time is measured from the current time $t=0$, and
the 68\% confidence region on each pericentric passage is also shown.

\subsection{The `missing' trailing tail}

As to the fractions of luminous material shed at each pericentric passage, it is unfortunately difficult to 
transform the inferences of the stream model into dimensional units. This would require resolved colour information
along the stream. On a qualitative level, our modelling approach measures that the fractions
of material shed at each pericenter decrease with time, and that the first interaction with the host strips the progenitor 
of a significant amount of luminous mass ($\approx0.6$ of the luminance of the total shed mass). 

Note however, that while material shed at the second and third pericentric passages is entirely probed by the available data (see Fig.~\ref{bestfitt}),
visible material shed around pericenter 1 is limited to the leading tail only. The corresponding trailing tail 
lies entirely outside the region probed by Fig.~\ref{rawdat}. Its location is illustrated in Fig.~\ref{distandvel}: the trailing tail extends to over 
100 kpc in projection and, as shown in the lower panel, its actual radial distance from the centre of NGC 1097 reaches
over 230 kpc, almost out to the nominal virial radius $r_{200}$. 
Our model assumes that equal amounts of material are lost from leading 
and trailing tidal radii. As shown by Fig.~\ref{distandvel}, under this assumption the trailing tail contains a significant amount of mass
and should be almost as bright as the other plumes. 

There are however at least two possible reasons as to why
the trailing tail might have escaped detection so far. The first is that the hypothesis of equal escape fractions at the leading and trailing
tidal radii is wrong for sufficiently close pericentric passages. Material could be lost in 
larger amounts from the leading tidal radius, where tides are stronger, so that our working assumption may well be over simplistic. 
The second possibility has to do with the exact profile of the density slope at large radii. Our model can account for significantly 
steep profiles, but does not allow for truncations at even larger radii. A density slope that continues to steepen with radius would
cause an additional increase in the radial distances reached by the trailing tail, up to determining the escape 
of part of such material.

\subsection{Line-of-sight kinematics}

\begin{figure}
\centering
\includegraphics[width=\columnwidth]{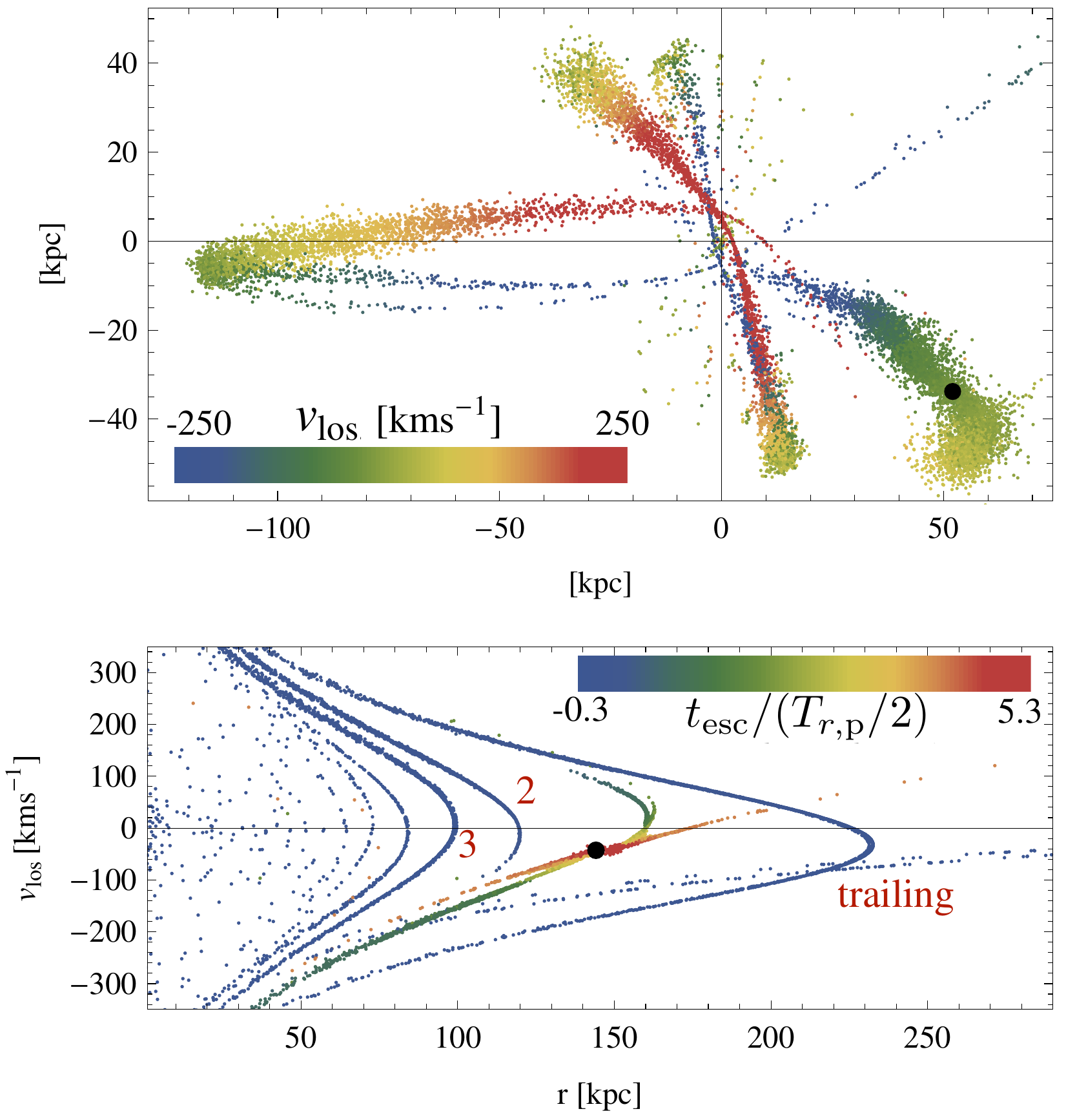}
\caption{{\it Upper panel:} a map of projected kinematics of the stream.
{\it Lower panel:} line-of-sight velocity against galactocentric distance; as in Fig.~\ref{bestfitt}, colour coding indicates the
shedding time, to facilitate the identification of the different plumes.
\label{distandvel}}
\end{figure}

Figure~\ref{distandvel} displays a map of the line-of-sight velocity of the stream material, 
with respect to the systematic velocity of NGC 1097. Note that the models are symmetric with respect to the transformation
\begin{equation}
x_{\rm los}\rightarrow -x_{\rm los}\ ,
\end{equation}
and additional information is required to fix the sign of the line-of-sight velocities, as well as the sign of the 
line-of-sight position of the progenitor at the current time $x_{\rm p, los}$. \citet{PG10} measure that the 
progenitor is receding faster than NGC 1097, and we fix the direction of the line-of-sight accordingly. In more detail, \citet{PG10} measure 
\begin{equation}
v_{\rm los}^{\rm obs}\approx-30\pm30\ {\rm kms}^{-1}\ ,
\label{vlosobs}
\end{equation}
while the predictions of the model are
\begin{equation}
\left\{
\begin{array}{rcll}
v_{\rm p, los}&= &-51^{+14}_{-17} &{\rm kms}^{-1}\\
x_{\rm p, los}&= &134\pm17 &{\rm kpc}
\end{array}
\right.  \ ,
\label{vlosmod}
\end{equation}
so that the progenitor is nearer to us than NGC 1097.
As shown in Fig.~\ref{distandvel}, the entire leg is found receding, with a clear velocity gradient along its length. 
An even more marked gradient can be observed along {\it plume 2}, while {\it plume 3 }has a high internal velocity dispersion 
due to containing at the same time material that approaches and recedes our viewpoint. This is also clear from 
the bottom panel of the same Figure, which shows a plot of radial distance against line-of-sight velocity. 
Colour coding is the same as in Fig.~\ref{bestfitt} and shows the escape time of the stream particles. 
The extended trailing tail has mainly positive line-of-sight velocities, as {\it plume 2}; {\it plume 3} instead is folded 
around its apocenter.

%

\subsection{Internal kinematics}

\begin{figure}
\centering
\includegraphics[width=\columnwidth]{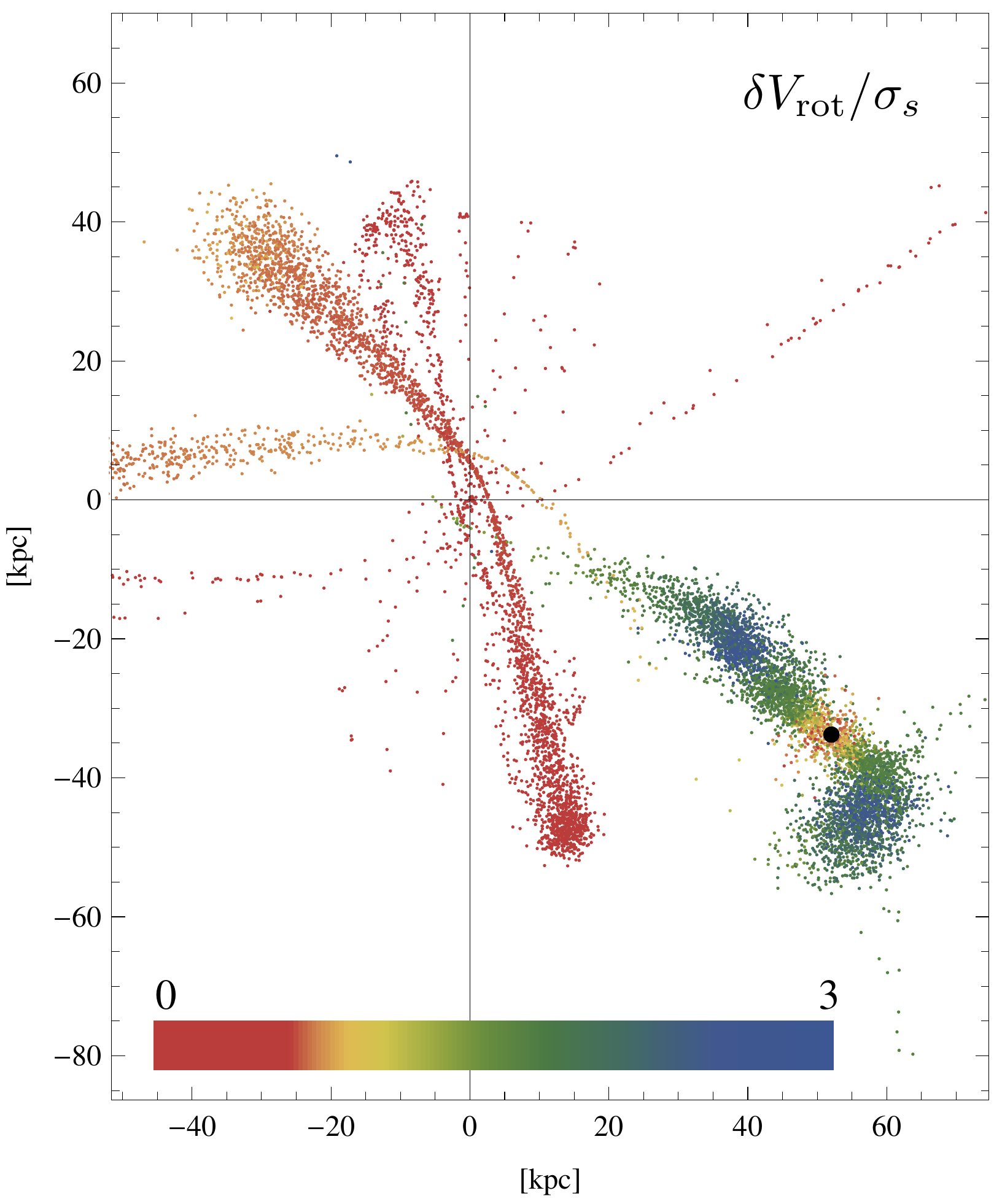}
\caption{The ratio between the ordered $\delta V_{\rm rot}$, and random $\sigma_{\rm s}$, components 
of the escape velocity of each stream member (colour coding is illustrated by the legend). Both 90 degree turn 
and puffy `knee' in {\it plume 1}, the `dog leg', are the signature of the internal angular momentum of the
progenitor. 
\label{escandrot}}
\end{figure}

We find that the random velocities at escape increase with time, from $\sigma_{\rm s}\approx 1$ kms$^{-1}$ for material lost 
at the first pericentric passage, to $\sigma_{\rm s}\approx$ 3 kms$^{-1}$ at the current time.
More interestingly, the model finds that non negligible ordered kick velocities
are required to explain the morphology of the dog leg. These are larger than the random velocities and
therefore correspond to internal rotation in the progenitor. Fig.~\ref{escandrot} shows the 
ratio between the ordered and random components of the kick velocities at escape for all stream members.
The X-shape, defined by {\it plumes 2, 3} and {\it 4}, is well described by material escaping with negligible
coherent velocities. However, in order to reproduce the dog leg in {\it plume 1}, values of $\delta V_{\rm rot}/\sigma_{\rm s}\gtrsim 1$
are required. 

Opposite signs of the coherent escape velocities for trailing and leading material allow them to stream 
to opposite sides of the main track of the leg. The trailing part of such material entirely makes up for the peculiar 90 degree 
turn at the edge of {\it plume 1}, currently at apocenter (see also the bottom panel of Fig~\ref{distandvel}). 
Opposite kick velocities in the leading material cause the shift and peculiar morphology of the bright `knee' of the dog leg.
Both these features are the signature of internal rotation in the progenitor. The required rotation is prograde, and although limits on the precise inclination of the 
internal angular momentum are loose, we find that $\phi_{\rm inc}<-0.3$, i.e.
the disk is indeed somewhat inclined with respect to orbital plane.

As noted in Section~5, our model seems to produce a dog leg that is slightly puffier than required by the data, and somewhat 
shorter, as shown by the residual map of Fig.~\ref{bestfitt}. We advocate this is due to our certainly simplified 
treatment of the escape conditions, and in particular of the relation between escape velocities and 
spatial spreads in the initial conditions. The need for ordered kick velocities increases the value of
$\bm{\delta v}$, which in turns makes the stream locally thicker because of the prescriptions~(\ref{semiax}).

Finally, we can use our measurement of the remnant's mass at the current time, eqn.~(\ref{mnow})
to estimate the velocity dispersion of the remnant: 
\begin{equation}
\sigma_{\rm p}\approx 12^{+14}_{-6}\ {\rm kms}^{-1}\ ,
\label{vlosobs}
\end{equation}
where we have adopted the virial relation provided by \citet{NA11} and the half-light radius of the 
remnant as measured by \citet{PG10}, $r_{\rm h}\approx300$ pc.

%

\section{Discussion}

\subsection{Limits of the model}

In this Section, we discuss some of the limitations of our modelling approach, and the possible 
biases they may introduce. 

\subsubsection{Dynamical friction}
We start with dynamical friction, which is not taken into account by the model, so that
the parameters of the progenitor's orbit remain constant in time since infall. 
We first note that dynamical friction operates on the progenitor, but is negligible on the very light and diffuse material populating the stream once this is shed. 
For example, the morphology of the leading tail lost at the first pericenter --  plumes 2, 3, and 4 -- is 
unaffected by dynamical friction. As a consequence, the logarithmic slope of the total density profile of the host, 
which mainly derives from the shape of such tail, is largely unaffected by our working hypothesis.

On the other hand, the orbital evolution of the progenitor driven by dynamical friction may cause shifts in the pericentric distances reached
at each successive pericentric passage. In turn, such shifts cause variations to the speed of the differential 
streaming of material shed around that time. In other words, while material lost at the second and third pericentric passages 
is assumed to be lost around the pericentric distance of the first passage, its actual distance might have been smaller due to 
dynamical friction. This translates in a bias on the inferred masses of the remnant at the second and third passage. 
As the pericenter shrinks, the speed of the differential streaming increases (A15). As a consequence, masses that are even lower than
in eqns.~(\ref{mperi2}-\ref{mnow}) would be required after the first pericentric passage -- or conversely an even higher initial mass~(\ref{mperi1}). 
Therefore, dynamical friction would make the mass evolution of the progenitor even starker. 

It turns out, however, that dynamical friction has little effect on the precise case in hand. 
We show this by providing an estimate of the relative loss of angular momentum caused by each
pericentric passage. We assume that the friction force can be described as in~\citet{BT87}, i.e. as it 
were excerpted by a sea of Maxwellian particles, with characteristic velocity $v_c$. As the orbit of the progenitor 
is strongly radial, we approximate the interaction time with $\delta t\approx 2 r_{\rm peri}/v_{\rm p}$, where 
$v_{\rm p}$ is the velocity of the progenitor at pericenter.
Under these assumptions we find
\begin{equation}
{{\delta J_{\rm p}}\over{J_{\rm p}}}\approx- 2 c \ln\Lambda {{Gm}\over {v_{\rm p}^2\; r_{\rm peri}}} \left({v_c\over v_{\rm p}}\right)^2 \ ,
\label{friction}
\end{equation}
where $\Lambda\approx r_{\rm peri}v_{\rm p}^2/Gm$ is the usual Coulomb logarithm, and the precise value of the dimensionless constant $c$
depends on the hypothesis made on the phase space distribution of the deflectors, $c\approx0.43$ for a Maxwellian distribution.
Even at the first pericentric passage, when the progenitor is at its most massive, we find a very limited effect, ${{|\delta J_{\rm p}|}/{J_{\rm p}}}\lessapprox 0.05$, 
comparable with the uncertainty of our measurement (see eqn.(\ref{jcirc})).
After the considerable mass loss of the first interaction, dynamical friction becomes negligible, with ${{|\delta J_{\rm p}|}/{J_{\rm p}}}\lessapprox 0.002$
at the second pericentric passage, and accordingly smaller thereafter.

%

\subsubsection{Host mass evolution}
Another limitation is connected to the assumption of a static host potential. 
If the mass $M(r)$ increases as a function of time, this implies an evolution in the orbital
energy of each streaming particle. For a stream like the one studied here, this effect is expected to be small. In a Milky 
Way sized halo, the mass within 100 kpc is already in place to within a few percent since redshift $z=1$ \citep{JD07, BK10}.
For our stream, infall is at $z\approx0.5$ implying, on average, a very small mass evolution 
within the extension of the observed plumes.

More in detail, a first effect of mass evolution is to make streams somewhat longer and thicker. The precise change in the energy of each
particle depends on its orbit $\lambda$, which couples it to the time dependence of the host mass evolution:
\begin{equation}
\delta(E)_{\rm m.ev.}=\int_{\lambda(r,t)} {\partial \Phi(r,t)\over \partial t} {\rm d}t\ .
\label{hmevol0}
\end{equation}
Close stream particles with similar orbits experience a similar energy evolution, but for a stream that is long enough to have 
at the same time members in opposite orbital phases, eqn.~(\ref{hmevol0}) introduces an additional energy spread.
The energy spread of the shed material fixes the speed of the stream growth (A15), so that, for the same values of the progenitor mass and 
internal velocity dispersion, mass evolution can make long enough streams even longer and thicker.  As a consequence, this could potentially cause an 
overestimate in the masses of the progenitor and in the random velocities at escape.

The relevant dimensionless ratio that quantifies the importance of such effect is therefore the ratio
between the additional energy spread introduced by mass evolution, $\sigma(E)_{\rm m.ev.}$,
and the characteristic energy difference (between tail and progenitor) that fixes the average tidal streaming speed $\delta E_{\rm t}$.
For the particular case in hand, the orbital phases of stars lost after the first pericenter have not grown different enough
from each other to result in an appreciable energy difference from mass evolution. Therefore, the only part of the stream that might be 
affected is the long leading tail lost at the first interaction. Even if we largely overestimate the spread in energy introduced by mass evolution with 
${\sigma(E)_{\rm m.ev.}/ {E_\lambda}}\approx  {\delta M/ M}$,
we obtain that the effect on the angular length of the stream is still very small, of the same order of $\delta M/ M$, i.e. a few percent.

A second effect of mass evolution has to do with the shape of the orbits: if $M(r)$ increases maintaining approximately its
spherical symmetry, the orbital angular momentum is conserved while energy decreases. This circularises the orbits
and determines a reduction of their apocentric distances with time, potentially affecting our measurement of the 
logarithmic density slope. This mainly derives from the ratios between the apocentric distances reached by the different plumes,
so that when such reduction is uniform over all particles, no bias is created.

Instead, cold dark matter halos evolve in an inside-out manner: at recent times their mass has evolved 
more slowly in their inner regions then in their outskirts \citep[e.g.][and references therein]{Zh09}. 
As a result, it is the apocenters of the most energetic 
particles that are the most affected by mass evolution. While the potential well experienced by particles 
more deeply embedded in the potential stays unchanged, the particles with higher energies experience 
a reduction of their apocenters. The qualitative net effect is therefore a bias towards measuring lower 
values of $\gamma_o$: for a fixed energy gradient along the stream, a less steep density profile can 
compensate for the smaller differences between the apocenters of the different plumes determined by 
mass evolution.

\subsubsection{Spherical symmetry}
Finally, a more worrying limitation is that the present model uses spherically symmetric 
host density distributions. Dark matter haloes have been shown to be substantially non-spherical, 
and haloes with virial masses similar to what we infer for NGC 1097 
typically have axes ratios distributed around $a/c\approx0.6$ and $b/c\approx0.8$ \citep[][and references therein]{Sc12}.
On the other hand, the density distribution of NGC 1097 is certainly non spherical in its central 
regions, which are likely dominated by its stellar disk and bar. 

The main effects of a non spherical  
potential are of two different kinds. First, as the equipotential surfaces are non spherical, 
apocentric distances are not constant in time, and vary with orbital phase. Second, as particles experience
a net torque, their instantaneous orbital plane is subject to a time dependent precession. 
If the non sphericity of the total density distribution is only due to the baryons, the quadrupole moment 
at the radii of the plumes is small, and so are these effects: the equipotentials at those distances are 
almost spherical \citep[see for example][]{NA10}, and the precession of the orbital plane is not big enough 
to become apparent in a stream that extends for only three orbital oscillations. 
However, if the density profile of the halo itself is non-spherical out to large radii, then these effects are substantial, 
and it may appear surprising that we are in fact capable of describing the stream almost down to the noise using a spherical model.

We have made tests using single orbits, similar to those described in Section~4, in which mock datasets have been constructed by 
solving the equations of motion for tracer particles in homeoidal density distributions \citep{Ch69}. These have initial conditions
fixed by our inferences for the current position and velocity of the remnant, and integrate its orbit backward and foreword in time
to mimic the four observed plumes. The trial density distributions we use are less prolate than suggested by 
cosmological simulations, with axes ratios $a/c\approx0.85$ and $b/c\approx0.9$, constant with radius, 
and oriented at random with respect to the present, instantaneous orbital plane. We find that use of our spherical library to fit for such orbits 
is uniformly unsuccessful. Over three full orbital oscillations, orbits in these triaxial potentials loose their planar nature 
to the point that reproducing all four plumes (in this case apocenters) with a planar orbit from our library 
is impossible. In particular, we can never achieve $\chi^2$ levels that are comparable with
those obtained in either our single orbit tests in spherical potentials, or in our stream fit.  

Although more detailed explorations are required, this suggests that, outside its central regions, the
total density distribution of NGC 1097 is not substantially non spherical. Observational constraints on
the shape of dark matter haloes are quite loose, and almost completely limited to modelling the Sagittarius stream
around the Milky Way \citep[e.g.][and references therein]{LM10, VB14}. More recently, significant expectation is building up
for the inferences obtained modelling the thin streams of globular clusters \citep[][]{AK15, AB15}. However, in comparative terms, 
these are limited to inner regions, where baryons are still important. On the other hand, it is widely accepted that dissipative 
dynamics and the lifecycle of baryons makes haloes rounder in cosmological hydrodynamical simulations \citep[e.g.,][]{SK04,VD08,Bu15}
when compared to dark-matter-only runs. Our result provides observational evidence in this direction.

%

\subsection{Autopsy of a massive halo}

With a virial mass of $\approx 2\times 10^{12} M_{\odot}$, we find NGC 1097 is not considerably more massive than the Milky Way.
CDM expectations for the concentration of such a halo are $c=8^{+2.5}_{-2}$, with the mentioned values indicating the 
25\% and 75\% quantiles \citep[][]{AL14, AM08}. NGC 1097 is in very good agreement with these figures.
In the stream's morphology, we can see the total density profile bending from an inner slope of $\gamma(r_{\rm peri})\approx 1.5$ around the 
pericenter, $r_{\rm peri}\approx 4$ kpc, and can clearly resolve its steepening. Interestingly, the circular velocity remains approximately flat
at radii that are even larger than the break radius of the total density profile, $r_0\approx37$ kpc. The very extended stream
allows us to map the profile of the total logarithmic slope at least until the orbital apocenter, $r_{\rm apo}\approx 150$ kpc, 
where we find $\gamma(r_{\rm apo})=3.9\pm0.5$ (68\% confidence region). To our knowledge, this represents the first statistical measurement of 
the outer density slope of a single extragalactic halo, i.e. on data that do not derive from stacking.

As to the rather steep outer slope, our result certainly calls for the analysis of a wider sample of galaxies,
so to establish whether NGC 1097 is just a case in the distribution of outer slopes, or whether the outer density profile 
of galactic dark matter haloes is systematically steeper than expected. We recall that cosmological simulations show that the 
outer regions of dark matter haloes should be characterised by substantial halo-to-halo variations. \citet{AR99} suggest 
that the outer slope of galaxy-sized haloes correlates with environment and even when selecting isolated haloes only, 
they measure that the 68\% confidence interval for the outer slope is sizeable, $2.5\leq\gamma_o\leq3.8$ \citep[see also][]{FP06}. 

Recently, 
\citet{BD14} have made clear that the simple NFW profile is often a poor fit to the outskirts of $\Lambda$CDM haloes, even in the 
region $0.5\lesssim r/r_{\rm 200,m}\lesssim1$. They show the significant halo-to-halo scatter in the outer density slope at $z=0$, 
with many galaxy-sized haloes having envelopes that extend all the way down to $\gamma(r_{200, m})\approx-5$. Their analysis concludes 
that the value of the density slope around the virial radius is systematically associated to the mass accretion rate of the halo: all 
across the mass scale, steeper outer density profiles correspond to higher rates of recent physical mass accretion. However, 
note that our apocentric distance is not large enough to explore the radial range studied in \citet{BD14}, 150 kpc $\approx 0.35\; r_{\rm 200, m}$,
underlying the need for a statistics of extended streams and measurements of more outer density slopes.

In fact, it is quite surprising that using a simple, spherical broken power law profile we can describe the stream almost down to the noise.
Apparently, the dynamics of the streaming material is not strongly affected by the details and local asperities of the gravitational potential, and 
a simple description like the one used here is sufficient to capture its salient properties. At the same time, our study shows that
a full exploration of the parameter space is absolutely necessary: \citet{HW03} concluded that X-shaped streams could not be 
created within spherically symmetric hosts because their suite of N-body simulations did not produce any. The approach of this paper
is considerably less numerically expensive than full N-body simulations, and therefore allows for a more systematic inspection of the wide
range of parameters. This is very promising and will allow the detailed study of more massive galaxies through the tidal features that encircle them. 
It certainly encourages the search for more low surface brightness, thin and extended streams. Here, we have made use of the projected morphology of the stream 
only, but kinematics of discrete tracers will be particularly helpful.

%

\subsection{The life of a cannibalised dwarf galaxy}

\citet{PG10} report that the remnant of the progenitor is highly symmetric, and has a well developed core-halo structure.
Its nucleus is bright ($\approx 5.9\times 10^5 L_{V,\odot}$) and compact, but its half-light radius, certainly smaller than 80 pc, could not
be measured precisely because of seeing. 
The halo of the remnant, instead, is observed to extend for the full width of the stream, for a luminosity of $\approx 2.6\times 10^6 L_{V,\odot}$
and an half-light radius of a few hundred parsecs. Its spectrum suggests its metallicity is not too low, 
with $[{\rm Fe}/{\rm H}]\approx -0.7$ providing a reasonable fit. 

Our analysis reconstructs that this is the remnant of a disky dwarf, and that the progenitor was considerably more massive at infall, 
with a total mass of $\approx 2.5\times10^{10} M_{\odot}$ within a few kpc (see eqn.~(\ref{mperi1})). Infall happened 
$\approx 5.4$ Gyr ago, and in the mean time, the progenitor has experienced three pericentric passages.
The first has been the most destructive, reducing its total mass by over 1.5 orders of magnitude. Most likely,
this corresponds to the removal of the largest part of its extended dark matter halo, leaving the remnant with only a few percent
of its initial mass after just one interaction. The progenitor was internally rotating, as testified by the peculiar morphology of the dog leg,
which requires ordered kick velocities $\delta V_{\rm rot}/\sigma_{s}>1$ for the material lost around the second pericentric passage. 

Although estimating the total luminosity of the stream is not an easy task,  
from its approximate shape and mean surface brightness, \citet{PG10} obtain $L_{\rm str}\approx 5 \times 10^7 L_{V,\odot}$. 
This implies that, during its post-infall evolution, the progenitor has also lost the largest part of its stellar component.
With a lower bound to the total stellar mass of $\approx 10^8 M_{\odot}$, and assuming that the virial mass of the progenitor at infall 
is at least $M_{\rm vir}\approx 10^{11}M_\odot$, such a dwarf galaxy is then roughly comparable with the expectation of most abundance 
matching recipes \citep[e.g.][]{Gu10,PB13}. Furthermore, our progenitor is also in rough agreement with the 
relation between nuclear stellar mass and total galaxy mass within the star's characteristic radius.
The relation constructed by \citet{LF06} using the galaxy population of the Virgo Cluster implies that 
a nuclear stellar mass of $\approx 2\times 10^6 M_\odot$ should correspond to a progenitor total mass 
of $\approx 2\times 10^9$ within the effective radius of its stellar distribution. 

Comparison between our measurement of the total mass of the remnant, eqn.~(\ref{mnow}), 
and the estimate of its stellar mass of $\approx 6\times 10^6 M_{\odot}$ obtained by \citet[][]{PG10} suggests that 
it may still contain some residual dark matter in the central regions. The current remnant is not massive enough to be 
subject to substantial tidal friction (see Section~7.1), and will keep evolving almost
undisturbed within the halo of NGC 1097. As $r_{\rm t}\approx r_{\rm h}$, its diffuse halo will be entirely lost, most probably within a few Gyrs, and 
the only remains of the progenitor will be its bare compact nucleus. 

If we use the the current luminosity of the nucleus and the same mass-to-light ratio inferred by \citet{PG10} for the remnant, 
we get that the Roche lobe of the nucleus has a size of about a hundred parsecs when at the orbital pericenter. This implies that it 
will survive for a long time, details depending on the precise value of its poorly constrained half-light radius, $r_{\rm h}\ll80$ pc.
Tidal stripping of galaxy nuclei has been shown to be one of the channels of formation of `bright' Ultra Compact Dwarf galaxies \citep{Pf13,AS14}.
In the case of NGC 1097, we are directly assisting to one of the stages in the formation of a low-luminosity UCD \citep[][]{JB11,No14}, so that
the progenitor of the deg leg stream proves that such evolutionary path also determines the formation of a low luminosity tail in the population
of non-globular stellar clusters. This study shows that analysis of the stream properties is capable of providing a one-to-one 
association between the observed remnant and the structure and nature of the progenitor dwarf galaxy, through the 
reconstruction of its post-infall evolution.

\section{Summary}

By modelling the two-dimensional surface brightness map of its stellar stream, we have reconstructed the 
accretion of a dwarf galaxy on to the massive disk galaxy NGC 1097. We have shown that modelling the 
projected map of the stream alone allows us to constrain at the same time the initial properties of the 
progenitor galaxy, its post-infall evolution, and the density profile of the host. 
We find that:

\begin{itemize}

\item{We can fully describe the properties of the stream using a spherical model for the total density
distribution of the host galaxy. We perform tests using single orbits in triaxial potentials and find that this 
would not be possible if the halo of NGC 1097 were to be substantially triaxial out to large radii (i.e. with constant
axis-ratios).}

\item{The progenitor dwarf galaxy is on a strongly radial orbit, so that the stream extends out to over 150 kpc, allowing us to map the 
logarithmic slope of the total density profile $\gamma$ of the host out to $\gtrsim0.6 r_{200}$. The shape of the stream implies that
the density slope bends from an inner value of $\gamma(r_{\rm peri}\approx4\;{\rm kpc})=1.5\pm0.15$, to a quite steep value of
$\gamma(r_{\rm apo}\approx150\;{\rm kpc})=3.9\pm0.5$. }

\item{We can clearly resolve this steepening with radius and find that the location of the break radius is in very good agreement
with the predictions of $\Lambda$CDM, so that NGC 1097 has a concentration of $c_{200}=6.7^{+2.4}_{-1.3}$ and a virial (total)
mass of $M_{200}=1.8^{+0.5}_{-0.4} \times 10^{12} M_{\odot}$. }

\item{Our model predicts that the remnant of the progenitor has a line-of-sight velocity of $v_{\rm p, los}=-51^{-17}_{+14}$ kms$^{-1}$
with respect to NGC 1097, which is in extremely good agreement with the measured value of $v_{\rm p, los}^{\rm obs}=-30\pm 30$ kms$^{-1}$ 
\citep{PG10}.}

\item{The progenitor dwarf galaxy, which is now observed as a nucleated dwarf with a total stellar luminosity of $\approx 3.3\times 10^7 L_{V,\odot}$ \citep{PG10},
has in fact lost more than two orders of magnitude in total mass since infall, $5.4\pm0.6$ Gyr ago, and experienced three pericentric passages.
 Our inference for its initial mass is $\log_{10}[m(<3.4\pm1 {\rm kpc})/ M_\odot]=10.35\pm0.25$.}
 
 \item{The peculiar morphology of the dog leg stream is due to the initial properties of such progenitor dwarf, which had a non negligible internal
angular momentum and was then likely a disky dwarf. Prograde rotation is mirrored in the coherent escape velocities that are necessary for
reproducing both `foot' (i.e. sharp 90 degree turn) and `puffy knee' in the dog-leg morphology, material lost around the second pericentric passage.}

\item{The remnant today is observed to have a marked `core-halo' structure. Our inference for the current total mass in the remnant
implies that the stellar `halo' is not compact enough to be resilient to tides and will be lost entirely. Conversely, the bright and compact nucleus is unaffected
by both tides and dynamical friction and will survive for a long time. }

\item{With a luminosity of $5.9\times 10^5 L_{V,\odot}$, our model shows explicitly that tidal stripping of dwarf galaxies down to their nuclei is one of the
mechanisms that brings to the formation of UCDs, in this case of what would be called a low-luminosity UCD, and can reconstruct the properties 
of its progenitor.}

\end{itemize}
 
These results encourage the observational effort towards the search for more extended, reasonably thin, stellar streams 
around massive galaxies. The combination of a wider sample of systems and the statistical modelling approach presented here 
will allow us to unveil the properties of both massive haloes and dwarf galaxies.

\acknowledgments

It is a pleasure to thank Aaron Romanowsky for constructive comments. 
The DARK Cosmology Centre is funded by the DNRF.

\end{document}